\documentclass[a4paper,12pt,floatfix,fleqn]{revtex4}

\usepackage{lscape}
\usepackage{floatflt}
\usepackage{amsmath}
\usepackage{latexsym}
\usepackage{fancyhdr}
\usepackage{longtable}
\usepackage{bm}
\usepackage{natbib}
\usepackage{graphicx}
\usepackage{dcolumn}
\usepackage[english]{babel}
\usepackage[dvips]{epsfig}

\begin{document}
\title{Turbulence characteristics of the B\"odewadt layer in a large enclosed rotor-stator system \\}
\author{Anthony Randriamampianina}
\email{randria@irphe.univ-mrs.fr, Tel.33 (4) 96.13.97.67}
\author{S\'ebastien Poncet}
\email{poncet@irphe.univ-mrs.fr, Tel.33 (4) 96.13.97.75}
\affiliation{Institut de Recherche sur les Ph\'enom\`enes Hors Equilibre\\
 UMR 6594 CNRS - Universit\'es d'Aix-Marseille I \& II\\
Technop\^ole Ch\^ateau-Gombert, 49 rue F. Joliot-Curie, BP 146. 13384
Marseille c\'edex 13 - FRANCE - Fax. 33 (4) 96 13 97 09}
\date{\today}

\begin{abstract}
A three-dimensional direct numerical simulation ($3D$ DNS) is
combined with a laboratory study to describe the turbulent flow in
an enclosed annular rotor-stator cavity characterized by a large
aspect ratio $G=(b-a)/h=18.32$ and a small radius ratio $a/b=0.152$,
where $a$ and $b$ are the inner and outer radii of the rotating disk
and $h$ is the interdisk spacing. The rotation rate $\Omega$ under
consideration is equivalent to the rotational Reynolds number
$Re=\Omega b^2/\nu=9.5 \times 10^4$, where $\nu$ is the kinematic
viscosity of the fluid. This corresponds to a value at which an
experiment carried out at the laboratory has shown that the stator
boundary layer is turbulent, whereas the rotor boundary layer is
still laminar. Comparisons of the 3D computed solution with velocity
measurements have given good agreement for the mean and turbulent
fields. The results enhance evidence of weak turbulence at this
Reynolds number, by comparing the turbulence properties with
available data in the literature \cite{lygr01}. An approximately
self-similar boundary layer behavior is observed along the stator
side. The reduction of the structural parameter $a_1$ under the
typical value $0.15$ and the variation in the wall-normal direction
of the different characteristic angles show that this boundary layer
is three-dimensional. A quadrant analysis \cite{Kang98} of
conditionally averaged velocities is performed to identify the
contributions of different events (ejections and sweeps) on the
Reynolds shear stress producing vortical structures. The asymmetries
observed in the conditionally averaged quadrant analysis are
dominated by Reynolds stress-producing events in this B\"odewadt
layer. Moreover, Case 1 vortices (with a positive wall induced
velocity) are found to be the major source of generation of special
strong events, in agreement with the conclusions of Lygren and
Andersson \cite{lygr01}.

\vspace{0.2cm} \textbf{Keywords}: rotor-stator flow,
 direct numerical simulation, LDA, three-dimensional turbulent boundary layer.
\end{abstract}

\maketitle

\newpage
\section{Introduction}
\indent An increasing interest in rotating disk flows has motivated
many studies over more than a century. They are indeed among few
examples of three-dimensional flows that in the laminar case are
described by exact solutions to the Navier-Stokes equations.
Moreover, besides its primary concern to many industrial
applications such as turbomachinery, the rotor-stator problem has
proved a fruitful means of studying turbulence in confined rotating
flows. This specific configuration is one of the simplest case where
rotation brings significant modifications to the turbulent field.
Rotating disk flows are also among the simplest flows where the
boundary layers are three-dimensional and they are therefore well
suited for studying the effects of mean-flow three-dimensionality on
the turbulence and its structure.

\subsection{Rotating disk flows}
\indent Batchelor \cite{batc51} solved the system of differential
equations relative to the stationary axisymmetric flow between two
infinite disks. He specified the formation of a non-viscous core in
solid body rotation, confined between the two boundary layers which
develop on the disks. In contrast, Stewartson \cite{Stew53} claimed
that the tangential velocity of the fluid can be zero everywhere
apart from the rotor boundary layer. Mellor et $al.$ \cite{Mel68}
discovered numerically the existence of a multiple class of
solutions showing that the two solutions advocated by Batchelor
\cite{batc51} and Stewartson \cite{Stew53} can be found from the
similarity solutions. Daily and Nece \cite{dail60} have carried out
a comprehensive theoretical and experimental study of sealed
rotor-stator disk flows. They pointed out the existence of four flow
regimes depending upon combination of the rotation speed and the
interdisk spacing. These correspond respectively to two laminar
regimes, denoted I and II, and two turbulent regimes, III and IV,
each characterized by either merged (I and III) or separated (II and
IV) boundary layers. In the latter, an inviscid rotating core
develops between the two boundary layers and rotates with a constant
angular velocity and a quasi zero radial velocity, following the
findings of Batchelor \cite{batc51}. They provided also an estimated
value for the local rotational Reynolds number at which turbulence
originates with separated boundary layers, $Re_r=\Omega r^2 /
\nu=1.5 \times 10^5$ ($r$ is the radial location) for aspect ratios
$G \leq 25$ \cite{owen89}. However, experiments have revealed that
transition to turbulence can appear at a lower value of the Reynolds
number within the stationary disk boundary layer (the B\"odewadt
layer), even though the flow remains laminar in the rotor boundary
layer (the Ekman or Von K\'arm\'an layer). Itoh et $al.$
\cite{itoh92} have provided detailed measurements of the flow
characteristics within the turbulent boundary layers for an enclosed
rotor-stator system with an aspect ratio $G=12.5$. They reported a
turbulent regime occurring earlier along the stator side at $Re_r
\simeq 8. \times 10^3$, while along the rotor side, turbulent flow
develops later for $3.6 \times 10^5 < Re_r < 6.4 \times 10^5$. They
concluded that the mean velocity distributions inside the respective
boundary layers were determined only by the local Reynolds number
$Re_r$. Cheah et $al.$ \cite{chea94} performed detailed measurements
of the turbulent flow field inside a rotor-stator system enclosed by
a stationary outer shroud with an aspect ratio $G=7.87$ and for a
rotational Reynolds number varying within the range $0.3 \times 10^6
\le Re \le 1.6 \times 10^6$. At the highest value of $Re$, they
found a laminar behavior of the Ekman boundary layer over the inner
half of the cavity and a turbulent behavior towards the outer radial
locations, which corresponds to $Re_r=4. \times 10^5$ for the
occurrence of turbulent flow along the rotor side. A different
behavior was reported for the B\"odewadt boundary layer, which is
turbulent at the lowest rotation rate considered. Differences in
turbulence characteristics were observed between the rotor and
stator sides and attributed to the effects of the radial convective
transport of turbulence. In stability experiments over a free
rotating disk, Wilkinson and Malik \cite{WILK85} found the
transition to turbulent flow to occur at the range $2.9 \times 10^5
\le Re_r \le 3.1 \times 10^5$. In his review of the
laminar-to-turbulent transition, Kobayashi \cite{koba94} reported
that the flow over a rotating disk remains laminar for values of the
local Reynolds number $Re_r \leq 4.5 \times 10^4$ and is fully
turbulent for $Re_r$ greater than about $3.9 \times 10^5$. Later,
Gauthier et $al.$ \cite{gaut99} have noticed for $G=20.92$ that
turbulence appears progressively toward the centre with spirals at
the periphery for $Re > 8.73 \times 10^4$. Even though their
geometry does not include any shaft, these values are close to the
ones used in the present study, assuming that the effect of the
shaft is less important for the occurrence of turbulence since the
flow remains always laminar near the axis. Schouveiler et $al.$
\cite{scho01} have identified two main routes for the transition to
turbulence according to the aspect ratio. For $G \leq 14.01$, the
boundary layers are separated and the transition was found to occur
through a sequence of supercritical bifurcations leading to wave
turbulence, resulting from the interaction between circular and
spiral rolls. For $G \geq 55.87$, the boundary layers are merged.
They observed the formation of localized turbulent structures in the
form of turbulent spots through subcritical transitions and
spatio-temporal intermittency.
\\
\indent Major experiments concerning the fully turbulent flow in a
shrouded rotor-stator cavity have been performed by Itoh et $al.$
\cite{IYIG90,itoh92} and recently by Poncet et $al.$
\cite{Pon04,PON05}. In the case of a closed cavity, Itoh et $al.$
\cite{IYIG90} measured the mean flow and all the Reynolds stress
components, and brought out the existence of a relaminarized region
towards the axis even at high rotation rates. When an inward
throughflow is superimposed, Poncet et $al.$ \cite{Pon04} showed,
analytically, that the entrainment coefficient $K$ of the fluid,
defined as the ratio between the tangential velocity in the core and
that of the disk at the same radius, can be linked to a local flow
rate coefficient according to a $5/7$ power-law, whose two
coefficients are determined experimentally. This law, which depends
only on the prerotation level of the fluid, is still valid as long
as the flow remains turbulent with separated boundary layers. Poncet
et $al.$ \cite{PON05} compared extensive pressure and velocity
measurements with numerical predictions based on an improved version
of the Reynolds stress modeling of Elena and Schiestel \cite{ES96}
for an enclosed cavity and also when an axial throughflow is
superimposed. In the case of an outward throughflow, they
characterized the transition between the Batchelor \cite{batc51} and
Stewartson \cite{Stew53} flow structures in function of the radial
location and of a modified Rossby number. All the comparisons
between measurements and predictions were found to be in excellent
agreement for the mean and turbulent fields.
\\
\indent Besides the theoretical or industrial aspects, turbulent
rotating disk flows are considered also as useful benchmarks for
numerical simulations because of the numerous complexities embodied
in this flow including wall effects, transition zone and
relaminarization. Most of the studies have been dedicated to
instability analyses in a shrouded cavity
\cite{serr01,JAC02,serr04}. Main contributions concerning turbulent
rotor-stator flows have been carried out by Lygren and Andersson
using DNS \cite{lygr01} and Large Eddy Simulation LES \cite{LYAN04}.
They have simulated the turbulent flow at $Re=4.6 \times 10^5$ in an
infinite disk configuration, using a restricted calculation domain:
($3.5 h$, $7 h$, $h$) according to the radial, tangential and axial
directions respectively. They have provided a detailed set of data
to analyse the coherent structures near the two disks \cite{lygr01}.
They also compared the results obtained from three LES models with
their DNS calculation \cite{LYAN04} and showed that the ``no model''
approach is the most effective. It suggests that improved subgrid
models have to be implemented to get closer agreement. A large
review of the main works concerning turbulent rotor-stator flows has
been performed by Owen and Rogers \cite{owen89} and Poncet
\cite{PONTHE}.

\subsection{Three-dimensional turbulent boundary layer}
\indent A three-dimensional turbulent boundary layer (3DTBL) is a
boundary layer where the mean velocity vector changes direction with
the distance from the wall, while the direction of the mean velocity
remains constant in a two-dimensional turbulent boundary layer
(2DTBL). Although the turbulence statistics and structures are
similar for 3DTBLs and 2DTBLs, there are some differences: the
vector formed by the turbulence stress is not aligned with the mean
strain rate in a 3DTBL. Another noticeable difference caused by the
three-dimensionality of the mean flow is the reduction of the
Townsend structural parameter $a_1$ (the ratio of the shear stress
vector magnitude to twice the turbulent kinetic energy $k$) below
the generally accepted value $0.15$ for conventional 2DTBLs
\cite{JOHN96}. The reader is referred to the work of Saric et $al.$
\cite{saric03} for a large review of the stability and transition of
three-dimensional boundary layers, in particular on swept wings and
rotating disks and to the one of Johnston and Flack \cite{JOHN96}
for a review of experimental studies and DNS of 3DTBLs (see also
Robinson \cite{ROB91}). 3DTBLs are usually found in engineering
flows such swept bumps \cite{WEBS96}, curved ducts, submarine hulls
or rotating systems. A particularly interesting feature in rotating
disk flows is that the boundary layer is three-dimensional from its
inception and leads to the appearance of characteristic strong
events. Indeed, the underlying structure does not result from
perturbing an initially two-dimensional flow but is inherent to a
boundary layer with a continuously applied crossflow.
\\
\indent From experimental investigations of turbulent pipe flow,
Corino and Brodkey \cite{cori69} observed the occurrence of bursting
events in the wall region. The burst begins with the acceleration of
a low-speed zone of fluid in both the sublayer and the buffer region
by a larger mass of fluid arriving from upstream. This is followed
by small-scale outward ejections of fluid from the low-speed region,
which interacts with the higher speed fluid to produce, at higher
distance from the wall, a chaotic motion bringing an increase in
turbulent mixing. Kim et $al.$ \cite{kim71} confirmed that nearly
all turbulence production occurs during the bursting process. Eaton
\cite{EAT95} showed in his review on experimental works of coherent
structures in 3DTBLs, that much attention has been focused on the
strength and symmetry of the vortices of opposite sign. Shizawa and
Eaton \cite{SHI92} used a generator vortex to embed a vortex in the
boundary layer approaching a wedge. The vortices decayed faster in
the three-dimensional boundary layer than in an equivalent
two-dimensional flow. Moreover, they found that Case 1 vortices
(having induced near-wall velocity in the direction of the crossflow
positive) produced weak ejections while the ejections from Case 2
vortices (with a negative wall induced velocity) were very strong.
\\
\indent Similar events have been observed later by Littell and Eaton
\cite{litte94} and Kang et $al.$ \cite{Kang98} from experimental
studies and by Wu and Squires \cite{wu2000} from LES of the
turbulent flow over a free rotating disk. The major experimental
work of the structural features of the 3DTBL over a rotating disk is
the one of Littell and Eaton \cite{litte94}, who investigated the
modification by the crossflow of the production of the shear stress.
Using the model of Robinson \cite{ROB91}, they found that the
crossflow leads to stronger ejections and weaker sweeps, and that
Case 1 vortices are the primary sources of generation of strong
ejections, while Case 2 vortices are responsible for most of the
strong sweeps. This is inferred from the presence of distinct
asymmetries of the vortices producing sweeps and ejections. It has
been confirmed numerically by Wu and Squires \cite{wu2000}. Chiang
and Eaton \cite{CHIA96} refined the previous work of Littell and
Eaton \cite{litte94} by hydrogen bubble visualizations. They
observed that Case 1 and Case 2 vortices were equally likely to
produce ejections but Case 1 vortices produce stronger ejections
than Case 2 vortices. Flack \cite{FLAC97} showed that
stress-producing events near the vortices in a curved bend were not
influenced by the sign of rotation of the vortices. Kang et $al.$
\cite{Kang98} revealed that Case 1 and Case 2 vortices were nearly
symmetric. They attributed the asymmetries to the changes in the
negative-Reynolds-shear-stress-producing events, which have less
relation to the streamwise vortical structures. More recently, Le et
$al.$ \cite{LE00} performed a simulation of two-dimensional flow
where the flow is suddenly set in motion and found that the imposed
three-dimensionality breaks up the symmetry and alignment of
near-wall structures. The results support some conclusions of
Littell and Eaton \cite{litte94}, when considering different values
of the wall-normal coordinate.
\\
\indent Compared to 3DTBLs over a rotating disk, where there are no
complicating effects arising from variations in geometry, 3DTBLs in
enclosed rotor-stator systems are more complex essentially because
of the confinement, which yields a dependance with the radial
location of all turbulence quantities. To avoid this difficulty,
Lygren and Andersson \cite{lygr01,LYAN04} considered using DNS
\cite{lygr01} and LES \cite{LYAN04} the fully turbulent flow in an
"infinite" rotor-stator system. They ascribed also the asymmetries
observed by Littell and Eaton \cite{litte94} to the coherent
structures but they concluded that Case 1 vortices are the primary
sources of generation of both strong ejections and strong sweeps.
\\
\indent The purpose of the present study is to shed light on the
turbulence characteristics of the three-dimensional boundary layer
developing along the stator wall in an actual configuration, where
complex effects from system parameters may influence the near-wall
structures caused by the mean three-dimensionality. An enclosed
rotor-stator cavity of large aspect ratio, which models a part of
the liquid hydrogen turbopump of the Vulcain engine (Ariane V), is
considered. The results are discussed in the context of the
idealized system proposed by Lygren and Andersson
\cite{lygr01,LYAN04}, who provided detailed data set for all
interesting quantities. Three main different aspects are to be noted
between the two studies: Lygren and Andersson \cite{lygr01} have
considered an infinite system with fully turbulent flow in merged
layers, while in the present case, the system is enclosed, with
separate boundary layers and laminar Ekman layer along the rotor
wall. The basic flow belongs to the Batchelor type family: the two
boundary layers are separated by a central inviscid rotating core
(regime IV \cite{dail60}). DNS calculations are compared with
velocity measurements to bring a better insight on the mean and
turbulent fields and to provide detailed data of the turbulent
boundary layer along the stator side as the boundary layer along the
rotor is laminar. The B\"odewadt layer is three-dimensional from its
inception with a continuously applied crossflow. In particular, a
quadrant analysis of the Reynolds shear stresses is performed to
show the contributions of various events occurring in the flow to
the turbulence production of vortical structures.

\section{Details of the experimental set-up}
\begin{figure}[!ht]
\begin{center}
\scalebox{1}{\includegraphics[width=11cm,height=5cm]{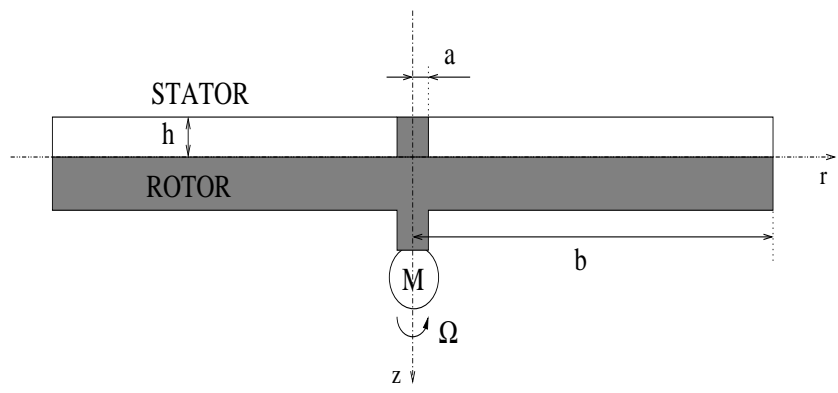}}
\caption{Randriamampianina et Poncet, Phys. Fluids.} \label{dispo}
\end{center}
\end{figure}

\indent The cavity sketched in figure \ref{dispo} is composed of a
smooth stationary disk (the stator) and a smooth rotating disk (the
rotor) delimited by an inner rotating cylinder (the hub) and an
outer stationary casing (the shroud). The rotor and the central hub
attached to it rotate at the same uniform angular velocity $\Omega$.
\\
\indent The mean flow is governed by three main control parameters:
the aspect ratio of the cavity $G$, the rotational Reynolds number
$Re$ based on the outer radius $b$ of the rotating disk and the
radius ratio defined as follows:

\begin{displaymath}
G=\frac{b-a}{h}=18.32 \qquad Re=\frac{\Omega b^{2}}{\nu}=9.5 \times
10^4 \qquad a/b=0.152
\end{displaymath}

\noindent where $\nu$ is the kinematic viscosity of water, $a=0.038$
m, $b=0.25$ m the inner and outer radii of the rotating disk,
$h=0.0116$ m the interdisk spacing and $\Omega$ the rotation rate of
the rotating disk. A small radial clearance $\delta=0.003$ m exists
between the rotor and the shroud ($\delta/b =0.012$).
\\
\indent A variable speed numerical controller drives the angular
velocity $\Omega$. The accuracy on the measurement of the angular
velocity is better than $1\%$. In order to avoid cavitation effects,
the cavity is maintained at rest at a pressure of $2$ bars.
Pressurization is ensured by a tank-buffer and is controlled by two
pressure gauges. The temperature is maintained constant using a heat
exchanger, which allows the removal of the heat produced by friction
in order to keep the kinematic viscosity $\nu$ of water constant.
\\
\indent The measurements are performed using a two component laser
Doppler anemometer (LDA). The LDA technique is used to measure from
above the stator the mean radial $V_r^*=V_r/(\Omega r)$ and
tangential $V_{\theta}^*=V_{\theta}/(\Omega r)$ velocities and the
associated Reynolds stress tensor components
$R_{rr}^*=\overline{v_{r}^{'2}}/(\Omega r)^2$,
$R_{r\theta}^*=\overline{v_{r}^{'} v_{\theta}^{'}}/(\Omega r)^2$,
$R_{\theta\theta}^*=\overline{v_{\theta}^{'2}}/(\Omega r)^2$ in a
vertical plane $(r,z)$. This method is based on the accurate
measurement of the Doppler shift of laser light scattered by small
particles (Optimage PIV Seeding Powder, $30$ $\mu$m) carried along
with the fluid. Its main qualities are its non intrusive nature and
its robustness. About $5000$ validated data are necessary to obtain
the statistical convergence of the velocity fluctuations
\cite{BAT76}.

\section{The numerical approach}

\subsection{Governing equations}
The motion is governed by the incompressible Navier-Stokes
equations. In a fixed stationary frame of reference, the
dimensionless momentum equations are:
\begin{eqnarray} \label {NSE1}
\frac {{\mathbf D}{\mathbf V}}{{\mathbf D}t} =
\frac{\partial{\mathbf V}}{\partial t} + \frac{1}{2}({\mathbf V}
\nabla{\mathbf V} + \nabla({\mathbf V}{\mathbf V})) = - \nabla {p} +
\frac{G(R_c+1)}{Re} \nabla^2 {\mathbf V}
\end{eqnarray}
\begin{eqnarray} \label {NSE2}
\nabla.{\mathbf V} = 0
\end{eqnarray}

\noindent where $\mathbf V$ is the velocity vector, $p$ the pressure
and $\nabla$ the nabla operator. We recall also that $Re$, $G$ and
$R_c$ are respectively the rotational Reynolds number, the aspect
ratio of the cavity and the curvature parameter defined by
$R_c=(a+b)/(b-a)=1.36$. The velocity and time scalings correspond to
$\Omega b$ and $h/(2b \Omega)$ respectively. In the meridional
plane, the space variables $(\overline{r},\overline{z})$ $\in$
$[a,b] \times [0,h]$ have been normalized into the square $[-1,1]
\times [-1,1]$, a prerequisite for the use of Chebyshev polynomials:
\begin{equation}
r = \frac{2\overline{r}}{b-a} - R_c, \,\,\,\,\,\,\,\,\,\,\,   z =
\frac{2\overline{z}}{h} -1.
\end{equation}

The `skew-symmetric' form proposed by Zang \cite{zang90} was chosen
for the convective terms in the momentum equations (\ref{NSE1}) to
ensure the conservation of kinetic energy, a necessary condition for
a simulation to be numerically stable in time.
\\
\indent The inner cylinder is attached to the rotor and so rotates
at the same angular velocity $\Omega$, while the other disk and the
outer cylinder are fixed. In order to maintain the spectral accuracy
of the solution, a regularization is introduced for the tangential
velocity component at the discontinuity between the rotating disk
and the stationary casing \cite{tave91,rand97,rand01}. In
Taylor-Couette flow problems, Tavener et $al.$ \cite{tave91}
mentioned that the effects of a clearance $\delta$ between the
rotating disk and the stationary casing on the flow patterns away
from the corners are negligible if $\delta$ remains sufficiently
small: $\delta/b < 0.02$. In the present case, the regularization
used in the numerical code is weak as well as in the experiment
$\delta/b=0.012$.

\subsection{Solution method}
A pseudospectral collocation-Chebyshev and Fourier method is
implemented. In the meridional $(r,z)$ plane, each dependent
variable is expanded in the approximation space $\mathcal{P}_{NM}$,
composed of Chebyshev polynomials of degrees less or equal than $N$
and $M$ respectively in the $r$ and $z$ directions, while Fourier
series are introduced in the azimuthal direction.

Thus, we have for each dependent variable $f$ :
\begin{equation}
 f_{NMK}(r,\theta,z,t) =
 \sum_{n=0}^{N}\,\sum_{m=0}^{M}\,\sum_{k=-K/2}^{K/2-1}\, \;{\hat f}_{nmk}(t) T_n(r)
 T_m(z) exp(ik \theta)
\end{equation}
where $T_n$ and $T_m$ are Chebyshev polynomials of degrees $n$ and $m$.

This approximation is applied at the collocation points, where the
differential equations are assumed to be satisfied exactly
\cite{canu87}. Since boundary layers are expected to develop along
the walls, we have considered the Chebyshev-Gauss-Lobatto
distribution, $r_i = cos(\frac {i \pi}{N})$ for $i \in [0,N]$ and
$z_j = cos(\frac {j \pi}{M})$ for $j \in [0,M]$, and an uniform
distribution in the azimuthal direction: $\theta_k = 2k \pi/K$ for
$k \in [0,K[$.
\\
\indent The time integration used is second order accurate and is
based on a combination of Adams-Bashforth and Backward
Differentiation Formula schemes, chosen for its good stability
properties \cite{vane86}. The solution method is the one developed and described in
\cite{hugth,rasp02}. It is based on an efficient projection scheme
to solve the coupling between velocity and pressure. This algorithm
ensures a divergence-free velocity field at each time step,
maintains the order of accuracy of the time scheme for each
dependent variable and does not require the use of staggered grids
\cite{hugu98}. A complete diagonalization of operators yields simple
matrix products for the solution of successive Helmholtz and Poisson
equations in Fourier space at each time step \cite{hald84}. The
computations of eigenvalues, eigenvectors and inversion of
corresponding matrices are done once during a preprocessing step.

\subsection{Computational details}
The spatial resolution corresponds to $N \times M \times K= 300
\times 80 \times 100$ in the radial, axial and azimuthal directions
respectively. The dimensionless time step was taken at $\delta t =
2.75 \times 10^{-3}$. The three-dimensional solution is obtained by
integrating the momentum equations, using an axisymmetric solution
as the initial condition into which a finite random perturbation is
introduced for the tangential velocity in each azimuthal plane.
After a statistically steady state was reached, turbulence
statistics were gathered during $15.09$ global time units in terms
of rotation period $\Omega^{-1}$. This is to be compared with the
time $2.9$ used by Lygren and Andersson \cite{lygr01} for fully
turbulent flows in both rotor and stator sides.

\section{Mean field and turbulence statistics}
We study the turbulent flow in a closed rotor-stator system of large
aspect ratio $G=18.32$. For the rotational Reynolds number $Re=9.5
\times 10^4$ considered here, the basic flow belongs to the regime
IV as defined by Daily and Nece \cite{dail60}: turbulent with
separated boundary layers known as a Batchelor flow structure. The
Ekman boundary layer on the rotor side and the B\"odewadt boundary
layer on the stator side are indeed separated by a central inviscid
rotating core. Note that the computed statistical data reported were
averaged in both time and in the homogeneous azimuthal direction.

\subsection{Mean field}
\begin{figure}[!ht]
\begin{center}
\scalebox{1}{\includegraphics[width=11.7cm]{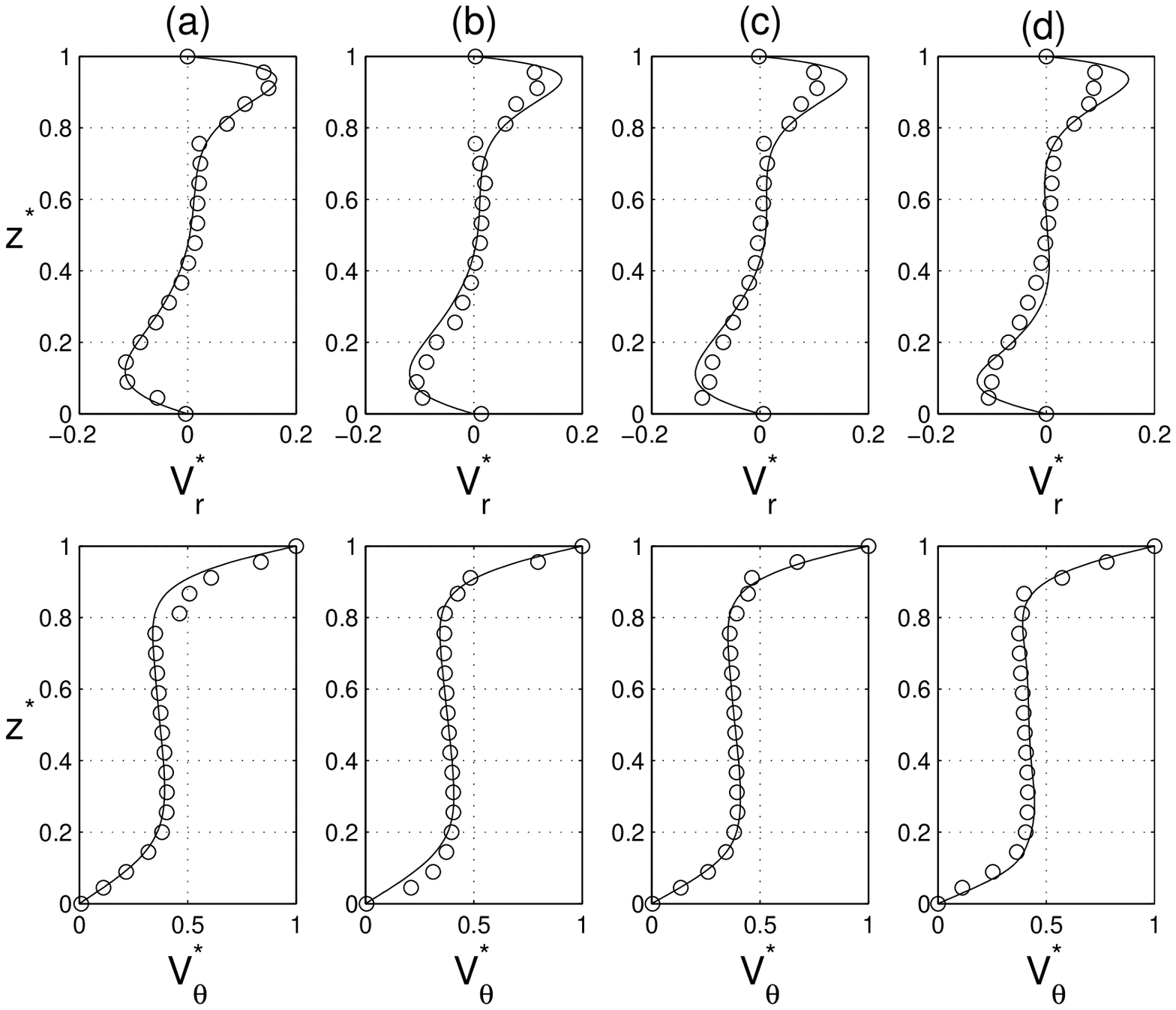}}
\caption{Randriamampianina et Poncet, Phys. Fluids.} \label{momoy}
\end{center}
\end{figure}

\begin{figure}[!ht]
\begin{center}
\scalebox{1}{\includegraphics[width=11.7cm]{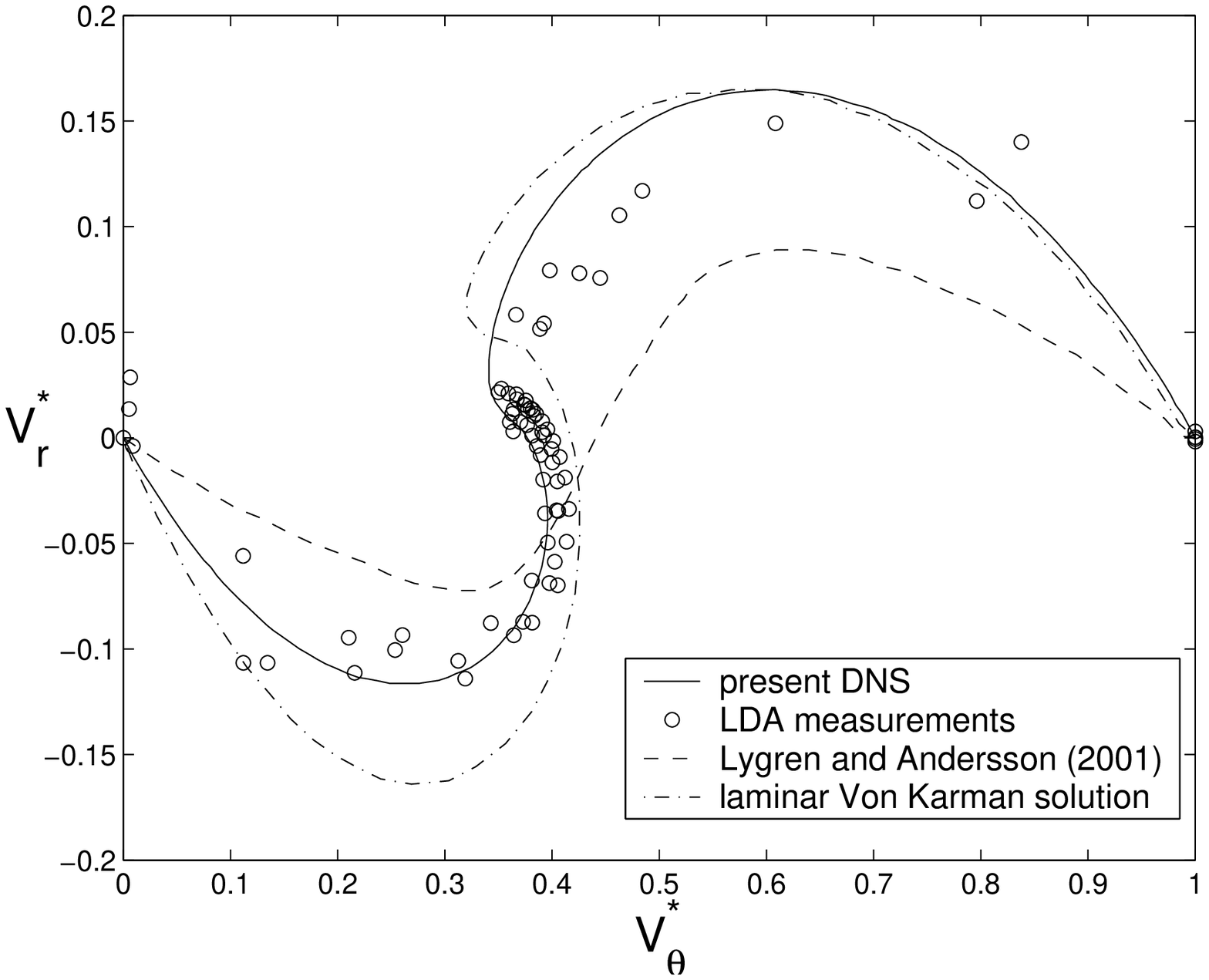}}
\caption{Randriamampianina et Poncet, Phys. Fluids.} \label{polar}
\end{center}
\end{figure}

The figure \ref{momoy} shows the axial profiles of the mean radial
$V_r^*=V_r / (\Omega r)$ and tangential $V_{\theta}^*=V_{\theta} /
(\Omega r)$ dimensionless velocities. The flow exhibits clearly a
typical Batchelor behavior \cite{batc51}, similar to the regime IV
defined in \cite{dail60}: two developed boundary layers on each
disk, separated by a central rotating inviscid core. The B\"odewadt
layer (towards $z^*=z/h=0$) along the stator side is centripetal
($V_r^*<0$). Its thickness, denoted $\delta_B$, is given by the
axial coordinate at which $V_{\theta}^*$ reaches $0.99 \times K$.
Note that $K$ is the entrainment coefficient of the rotating fluid,
defined as the ratio between the tangential velocity in the core and
that of the disk at the same radius. Then, the tangential velocity
$V_{\theta}^*$ ranges between $0$ and $0.99 \times K$ in that layer.
It is clearly shown in figures \ref{momoy}a to \ref{momoy}d that
$\delta_B$ decreases with the radius $r^*=r/b$. The Ekman layer
(towards $z^*=1$) is centrifugal ($V_r^*>0$) whatever the radial
location. Its thickness, denoted $\delta_E$, remains constant
independently of $r^*$, that is characteristic of laminar flows. In
that layer, the tangential velocity $V_{\theta}^*$ ranges between
$1$ and $1.1 \times K$. The rotating core is characterized by a
quasi zero radial velocity and by a constant tangential velocity.
The entrainment coefficient $K$ varies between $0.375$ and $0.418$
for the radial locations considered $0.44 \leq r^* \leq 0.8$, to be
compared with the theoretical value $0.431$ of Owen and Rogers
\cite{owen89} and to the semi-empirical value of $0.438$ for fully
turbulent flows proposed by Poncet et $al.$ \cite{Pon04}.
\\
\indent We report, in figure \ref{polar}, a polar plot of the mean
radial and tangential velocity components. The profile resembles
very well the ones reported by Itoh et $al.$ \cite{itoh92} from
their measurements. The polar profile in the B\"odewadt layer falls
between the typical fully turbulent behavior presented by Lygren and
Andersson \cite{lygr01}, which exhibits the characteristic
triangular form found in a three-dimensional turbulent boundary
layer, and the laminar solution obtained from the Von K\'arm\'an
\cite{VKT21} similarity equations. The polar profile in the Ekman
layer is closer to the laminar solution. This suggets that the flow
corresponds to a weakly turbulent flow, with turbulence mainly
prevailing along the stator side.
\\
\indent The 3D computed results are found here to be in close
agreement with the experimental data for the mean field.

\subsection{Turbulence field}
\begin{figure}[!ht]
\begin{center}
\scalebox{1}{\includegraphics[width=14cm]{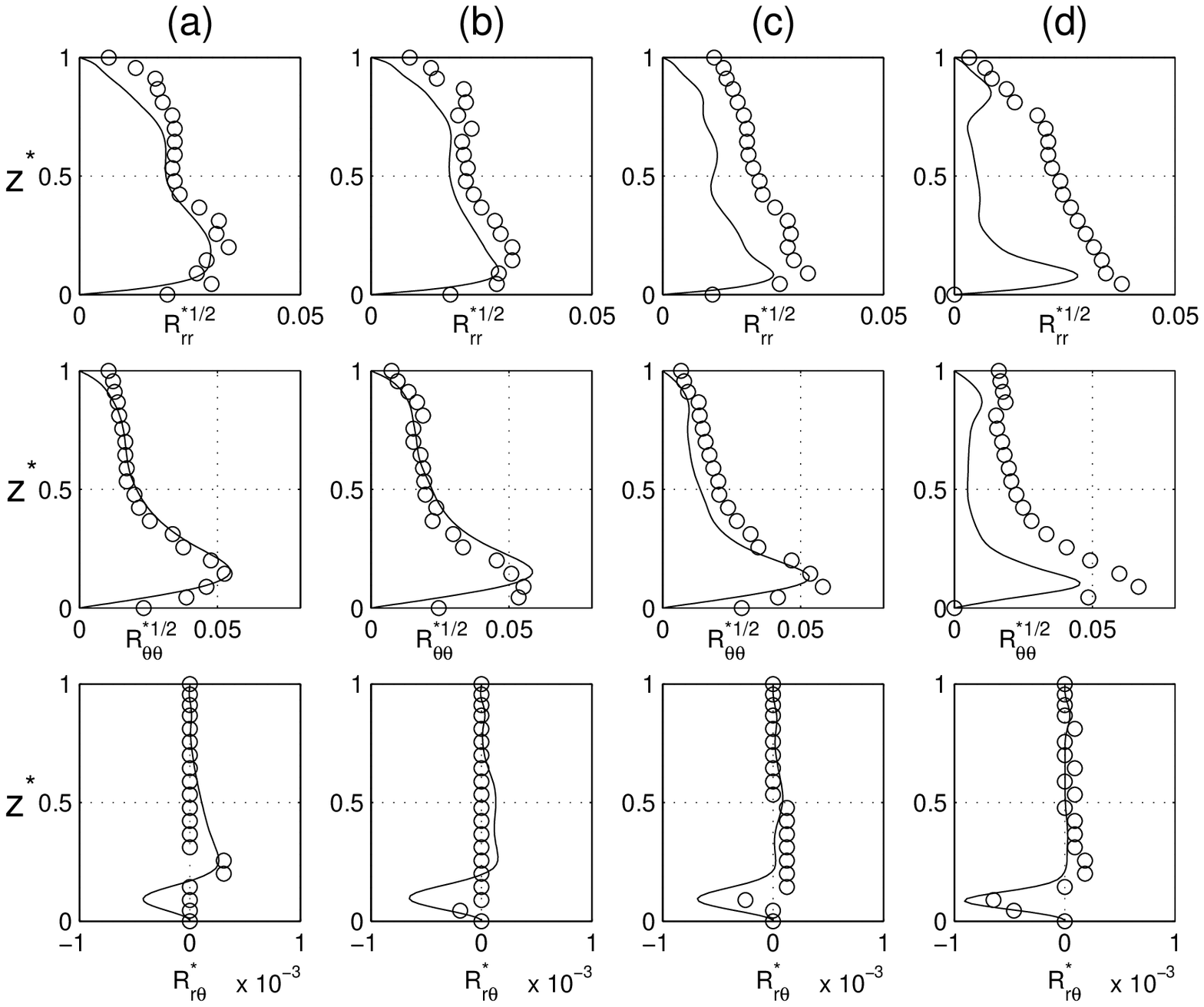}}
\caption{Randriamampianina et Poncet, Phys. Fluids.} \label{tuturb}
\end{center}
\end{figure}

\begin{figure}
\centering
\begin{picture}(280,180)
\put(-5,165){\shortstack{\large $\sqrt{\overline{v'^2_{\theta}}}$}}
\put(190,165){\shortstack{\large $\overline{v'_{\theta}v'_z}$}}
\put(-10,3){\includegraphics[width=6cm]{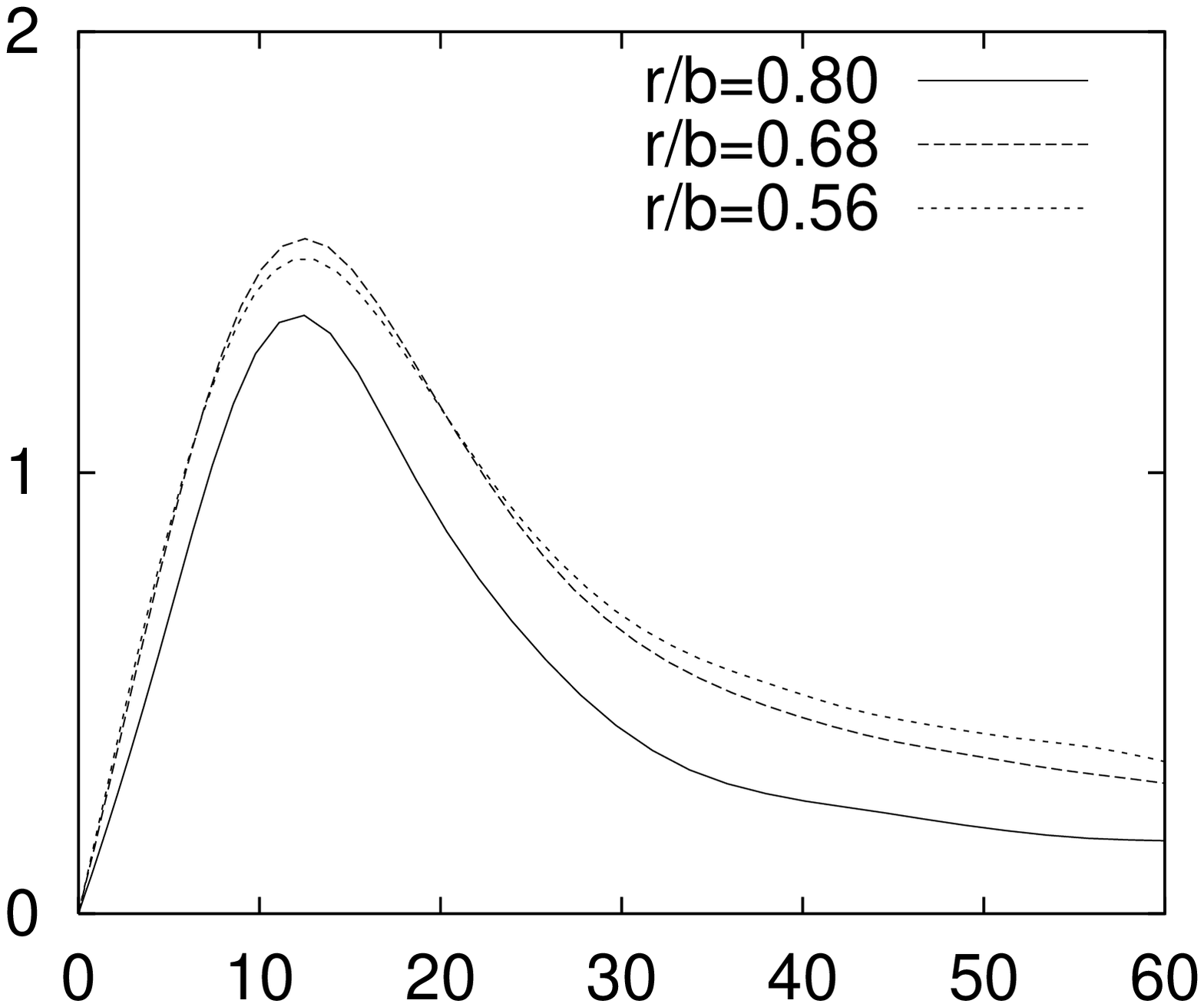}
\includegraphics[width=6cm]{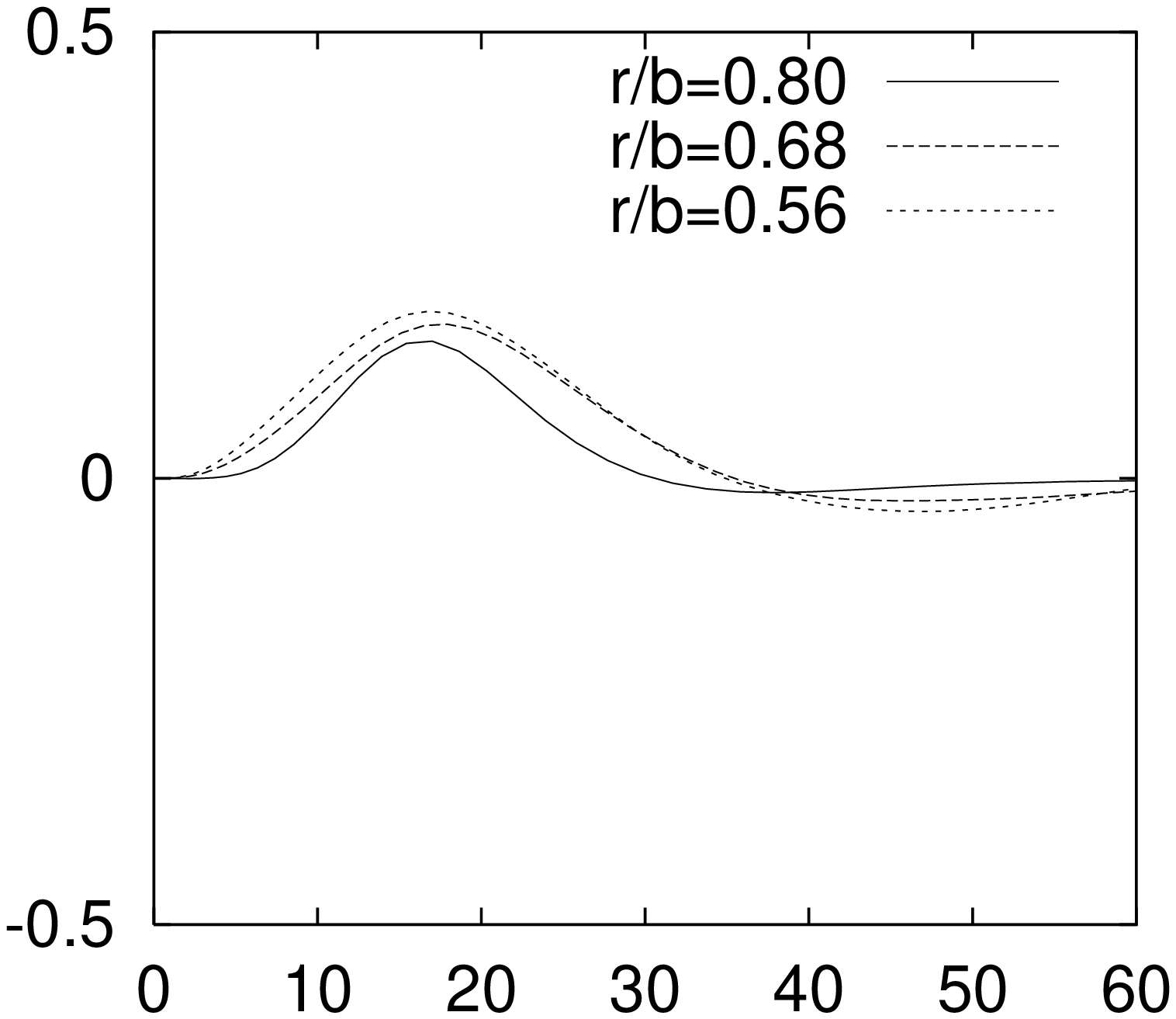}}
\end{picture}
\begin{picture}(280,180)
\put(-5,165){\shortstack{\large $\sqrt{\overline{v'^2_{r}}}$}}
\put(190,165){\shortstack{\large $\overline{v'_{r}v'_z}$}}
\put(-10,3){\includegraphics[width=6cm]{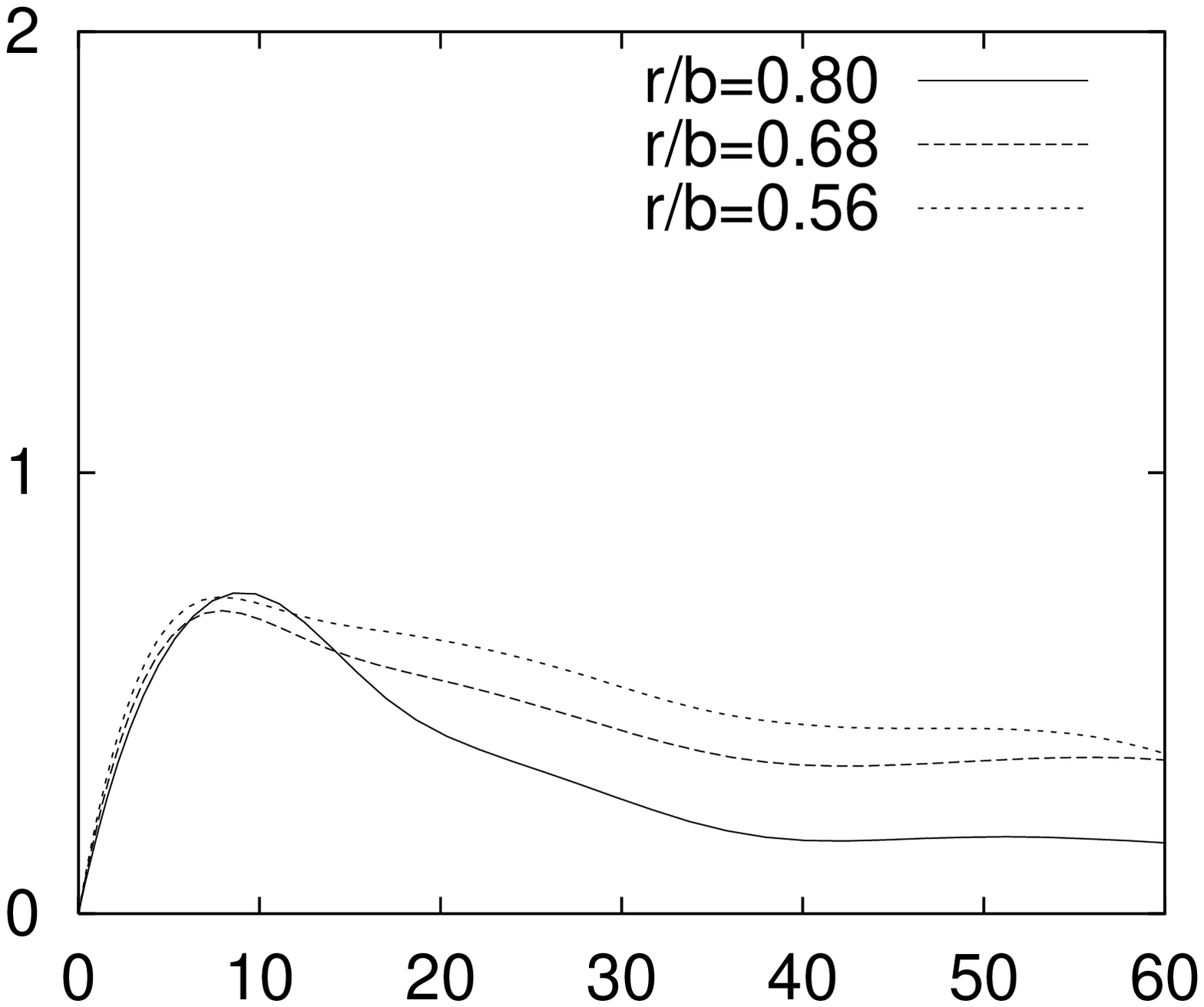}
\includegraphics[width=6cm]{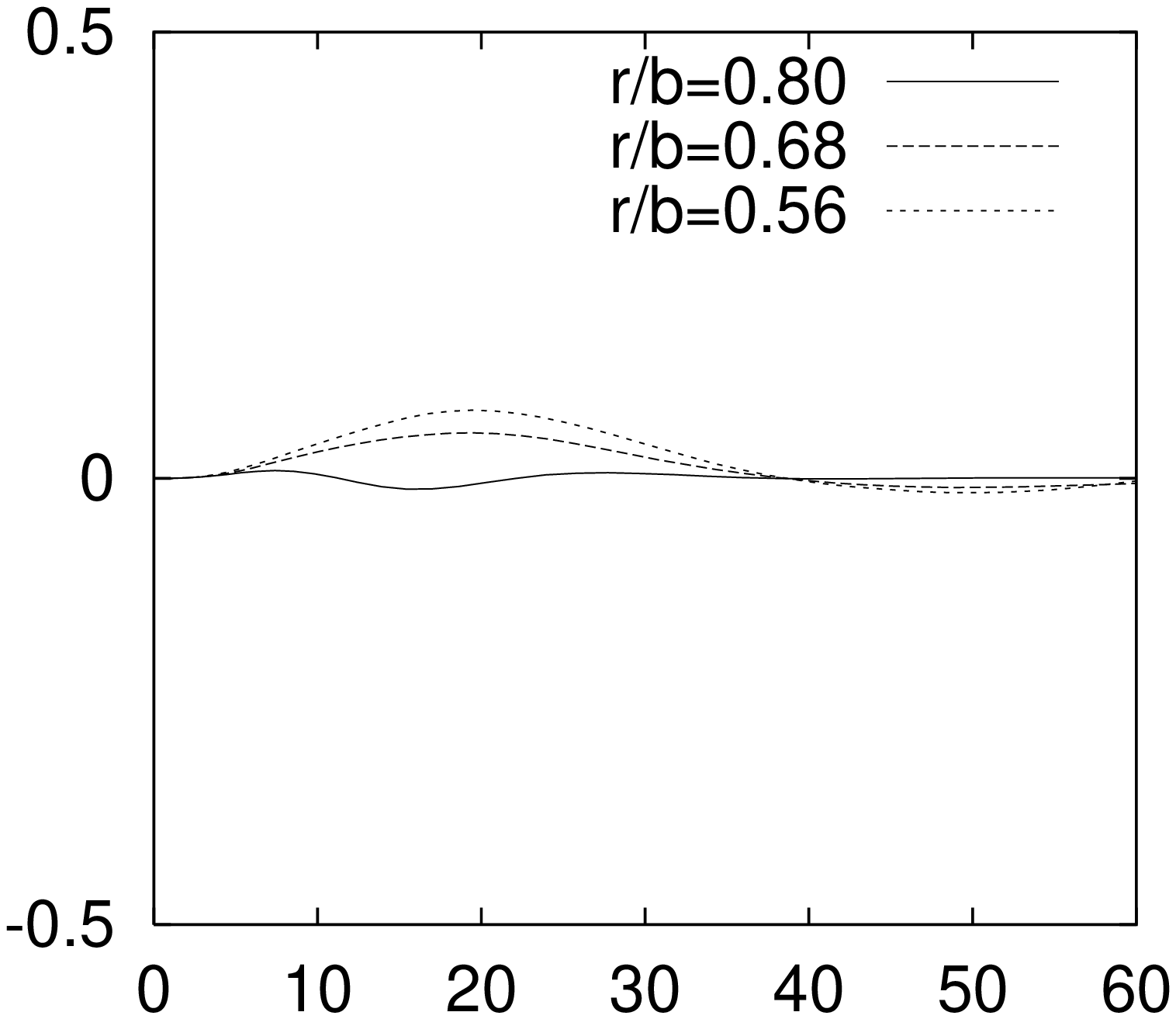}}
\end{picture}
\begin{picture}(280,180)
\put(-5,165){\shortstack{\large $\sqrt{\overline{v'^2_{z}}}$}}
\put(190,165){\shortstack{\large $\overline{v'_rv'_{\theta}}$}}
\put(-10,3){\includegraphics[width=6cm]{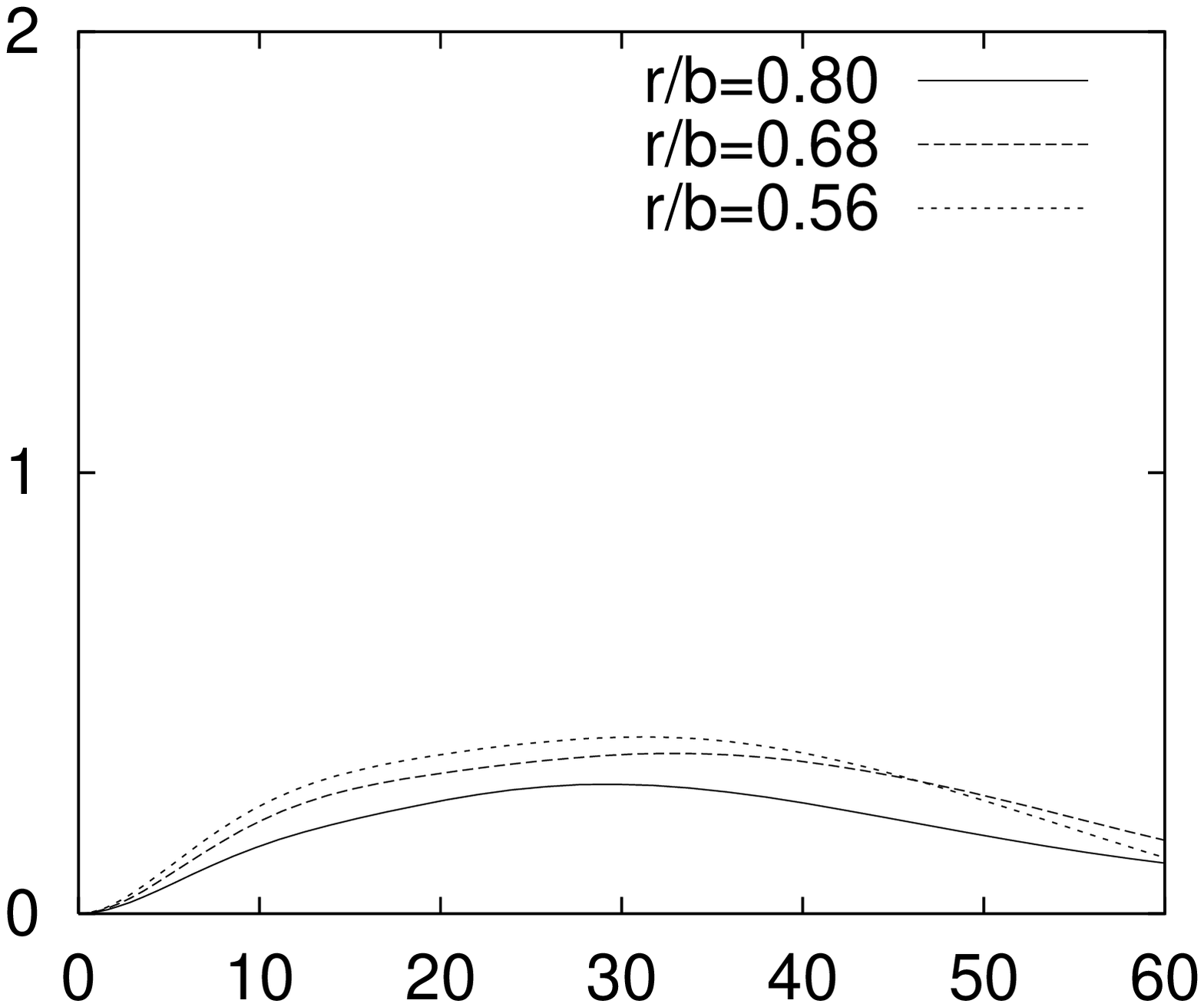}
\includegraphics[width=6cm]{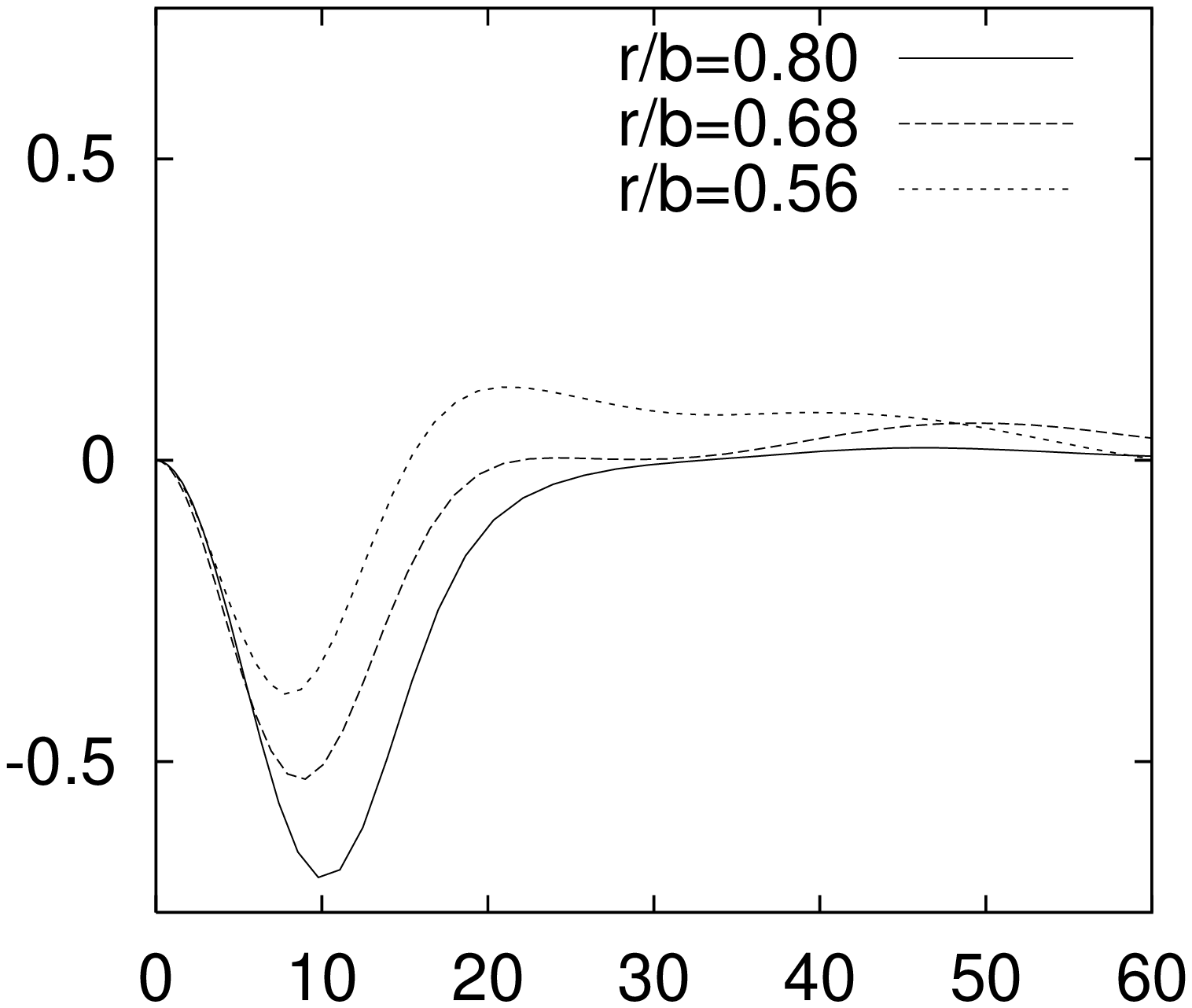}} \put(70,0){{\large
$z^+$}} \put(260,0){{\large $z^+$}}
\end{picture}
\caption{Randriamampianina et Poncet, Phys. Fluids.}
\label{Reystress}
\end{figure}

Comparisons between measurements and computations of the axial
variations of three components of the Reynolds stresses are
presented in figure \ref{tuturb} at four radial locations. Note
that, for the same flow control parameters, Poncet et
Randriamampianina \cite{Pon05cras} provided the corresponding
axisymmetric calculation and a laminar behavior of the flow is
obtained in the whole cavity. The 3D simulation presents behaviors
in good agreement with the measured data, even if the turbulence
intensities are rather weak. The axial profiles of $R_{r\theta}^*$
show that this Reynolds shear stress component is close to zero
everywhere except close to the stator wall. That means that there is
practically no turbulent shear stress at this rotation rate. The
turbulence intensities are mostly concentrated within the B\"odewadt
boundary layer, whereas the Ekman layer remains laminar. It is
noticeable that the flow along the stator becomes more turbulent
when one approaches the periphery of the cavity (for increasing
values of the local Reynolds number $Re_r$) considering the
experimental data. According to Cheah et $al.$ \cite{chea94}, on the
rotor side, the fluid is arriving from smaller radii where the flow
is laminar, while on the stator side, the fluid comes from larger
radii where turbulence prevails. In contrast, Itoh et $al.$
\cite{itoh92} mentioned that the turbulent flow in the stator
boundary layers is attributed to the unstable flow due to the
deceleration of the fluid.
\\
\indent Although the profiles from the simulation resemble the
behavior obtained from velocity measurements for $R_{rr}^*$ and
$R_{\theta\theta}^*$ until $r^*=0.68$, the DNS underestimates these
two orthogonal Reynolds stress tensor components towards the
periphery of the cavity. This discrepancy may result from the
different closures at the junction between the rotating disk and the
stationary outer casing. A small clearance equal to $\delta/b=0.012$
is present in the experimental rig which may generate some ingress
of fluid, while a regularized profile is imposed in the numerical
approach.
\\
\indent With the aim of providing a complete data set for the
enclosed system similar to the results reported in \cite{lygr01}, we
present in figure \ref{Reystress} the axial variation of the six
Reynolds stress tensor components near the stator wall as a function
of the wall coordinate $z^+ = z v_{\tau}/\nu$. To show the levels of
the present turbulence intensities compared to the fully turbulent
flow reported in \cite{lygr01}, the Reynolds stresses have been
normalized with the total friction velocity $v_{\tau}$, defined as
$v_{\tau}=(v^4_{\theta \tau}+v^4_{r \tau})^{1/4}$ with $v_{\theta
\tau}=(\nu \partial V_{\theta}/\partial z)^{1/2}$ the tangential
friction velocity and $v_{r \tau}=(\nu \partial V_{r}/\partial
z)^{1/2}$ the radial friction velocity. We recall that in
\cite{lygr01}, the geometry corresponds to two infinite disks, while
in the present case, confinement leads to radial variations of the
turbulent quantities, in particular $z^+$, as well as the B\"odewadt
layer thickness. Moreover, Lygren and Andersson \cite{lygr01} have
carried out computations for a fully turbulent flow in both the
rotor and stator sides, while turbulent behavior is only observed
along the stator wall with laminar flow towards the rotor wall in
our configuration. We note similar behaviors of the Reynolds stress
tensor components near the stationary disk between the two
simulations but with different levels. In particular, the normal
stress component along the axial direction $\overline{v'^{2}_{z}}$
is very weak, as well as the cross component $\overline{v'_r v'_z}$
in the $(r,z)$ plane.  As mentioned earlier, the boundary layer
thickness decreases with increasing radius towards the outer
stationary casing. It is noteworthy to recall that $v_{\tau}$ is an
increasing function of the radius as turbulence intensities become
higher towards the periphery. The variations obtained at different
radial locations fall more or less within the same profile,
suggesting an approximately self-similar boundary layer behavior for
this range of radial locations $0.56 \le r^* \le 0.8$.
\\
\indent The variation with the wall coordinate $z^+ = zv_{\tau}/\nu$
of the magnitude $\tau = (\overline{v'_{\theta} v'_z}^2 +
\overline{v'_{r} v'_z}^2)^{1/2}$ of the shear stress vector in
planes parallel to the disks is displayed in figure \ref{taut} at
three radial locations along the stator side. Also shown is the
variation of the magnitude $\tau_{tot}$ of the total shear stress
vector $(\nu \partial {V_{\theta}}/\partial z -
\overline{v'_{\theta}v'_z}, \nu
\partial{V_{r}}/\partial z - \overline{v'_{r}v'_z})$. These shear
stresses have been normalized by the total friction velocity
$v_{\tau}$, which varies with the radius. Unlike the findings of
\cite{lygr01} for infinite disk flow, after reaching a maximum,
$\tau$ decreases outside the boundary layer, while $\tau_{tot}$
decreases from its maximum value of $1$ within the boundary layer.
We can notice the low levels of the turbulent shear stress $\tau$,
with the main contribution from $\overline{v'_{\theta} v'_z}$ as
seen from the Reynolds shear stress components presented in figure
\ref{Reystress}. However, the profiles suggest again an
approximately self-similar boundary layer.
\\
\indent We display the isocontours of the turbulent Reynolds number
$Re_t=k^2/(\nu \epsilon)$ in figure \ref{kRet}a. As expected, high
levels of turbulence intensity are localized along the stator wall
with a maximum towards the junction between the stationary disk and
the outer casing. The presence of iso-contours close to the junction
between the rotating disk and the stationary outer casing suggests
that turbulence may start to develop at this zone. However, as
already observed from the Reynolds stress components, the low value
of the maximum of $Re_t=20.52$ confirms the weakly turbulent nature
of this flow. This maximum is to be compared with the maximum value
$Re_t=352$ obtained by Poncet et $al.$ \cite{PON05} for $Re=10^6$
and $G=23.89$. This weak level of turbulence is also confirmed by
the map of the turbulent kinetic energy (fig.\ref{kRet}b), which is
confined within the B\"odewadt boundary layer.
\\
\indent One characteristics of the three-dimensional turbulent
boundary layer is the reduction of the Townsend structural parameter
$a_1 = \tau/2k$, defined as the ratio of the shear stress vector
magnitude to twice the turbulent kinetic energy $k$. We have
reported in figure \ref{town} the variation at four radial locations
of $a_1$ versus $z/\delta_B$ at the stator side, where $\delta_B$ is
the B\"odewadt layer thickness (also function of radius). We can see
clearly a significant reduction below the limiting value $0.15$ for
a two-dimensional turbulent boundary layer, with behaviors similar
to those reported by Itoh et $al.$ \cite{itoh92} and Littell and
Eaton \cite{litte94} from their measurements. It confirms the
three-dimensional turbulent nature of the flow along the stator wall
\cite{lygr01,litte94}. This reduction of $a_1$ indicates that the
shear stress in this type of flow is less efficient in extracting
turbulence energy from the mean field. Moreover, it suggests that
irrotational inviscid motions dominate the outer region of the
B\"odewadt layer. However, even though this parameter is small, a
quadrant analysis will show that conditionally averaged velocities
can lead to a very strong contribution of the resulting shear stress
to the turbulence production, as detailed in the following sections.
\\
\indent To fix the three-dimensional nature of the B\"odewadt layer,
we display in figure \ref{gamm} the axial variation of the three
characteristic angles: the mean velocity angle $\gamma_m =
arctan(V_r/V_{\theta})$, the mean gradient velocity angle $\gamma_g
= arctan \left( \frac{\partial V_r/ \partial z} {\partial
V_{\theta}/\partial z} \right)$ and the turbulent shear stress angle
$\gamma_{\tau} = arctan(\overline{v'_r v'_z}/\overline{v'_{\theta}
v'_z})$. The profile of $\gamma_m$ clearly shows the continuous
change of direction of the mean velocity vector with the distance
from the wall, one of the major characteristics of three-dimensional
turbulent boundary layer. The angle remains in the range
$-45^{\circ} < \gamma_m \le 0^{\circ}$ within the boundary layer.
Another feature of 3DTBL is that the direction of the Reynolds shear
stress vector in planes parallel with the wall is not aligned with
the mean velocity gradient vector. Such a misalignment is observed
in the present simulation, with $\gamma_g$ smaller than
$\gamma_{\tau}$ near the disk and larger for $z/\delta_B \ge 0.55$,
as also mentioned by Lygren and Andersson \cite{lygr01}. However,
the lag between $\gamma_{\tau}$ and $\gamma_g$ is large towards the
extremities of the boundary layer with a maximum value about
$60^{\circ}$ to be compared with the value $18^{\circ}$ reported by
Lygren and Andersson \cite{lygr01} in infinite disk system. In their
numerical study of non-stationary 3DTBL, Coleman et $al.$
\cite{cole2000} obtained large values of the lag especially near the
wall, and inferred it from the slow growth of the 'spanwise'
component of the shear stress. These authors observed also the
change of the sign of the gradient angle. Such large values of this
lag make the assumption of eddy-viscosity isotropy to fail for the
prediction of such flows. In the present case, this feature
indicates a strong three-dimensionality with highly distorted flow
field resulting from the shear induced by rotation over the stator
wall, adding another complexity in comparison with the idealized
configuration in Lygren and Andersson \cite{lygr01}.

\subsection{Turbulence kinetic energy budgets}
The balance equation for the turbulent kinetic energy writes:
\begin{eqnarray} \label{TKEB}
A = P + D^T + D^{\nu} + \Pi - \epsilon
\end{eqnarray}

\noindent with the advection term $A=\overline{V_j}k_{,j}$, the
production term $P=-R_{ij}\overline{V_{i,j}}$, the diffusion due to
turbulent transport
$D^T=-\frac{1}{2}\overline{(v'_mv'_jv'_j)_{,m}}$, the viscous
diffusion $D^{\nu}=\nu k_{,jj}$, the velocity-pressure-gradient
correlation $\Pi=-\frac{1}{\rho}\overline{(v'_jp')_{,j}}$ and the
dissipation term $\epsilon=\nu\overline{v'_{j,m}v'_{j,m}}$. These
different terms are detailed in the Appendix.
\\
\indent Figures \ref{bilankz}a to \ref{bilankz}c show the axial
variations of the different terms involved in the transport equation
(\ref{TKEB}) at three radial locations. It is clearly seen that all
these terms vanish towards the rotor side, confirming the laminar
nature of this zone up to the stator boundary layer. At the stator
wall, the viscous diffusion balances the dissipation. Within the
B\"odewadt layer, even though some interaction between the different
terms involved is observed, the major contributions come from the
production, the dissipation and the viscous diffusion terms. The
production is balanced by the dissipation and the viscous diffusion,
which level increases at high radius in association with the
thickening of the boundary layer towards the periphery. The
production increases with increasing radius as already observed with
the levels of the normal Reynolds stresses (fig.\ref{tuturb}). The
maximum of the production term is obtained at $z^+=12$ for
$r^*=0.56$ and at $z^+=12.5$ for the two other radial locations,
which confirms the approximately self-similar behavior of the
B\"odewadt layer. These values are close to the value $z^+ \simeq
10$ reported by Willmarth and Lu \cite{WILL72} from experimental
studies in turbulent plane flow. The levels of the viscous diffusion
increase when moving towards the outer casing, where the highest
turbulence intensities prevail. At $r^*=0.8$ (fig.\ref{bilankz}c),
this increase is associated with a decrease of the dissipation term
and indicates that viscous effects still play an important role in
the turbulence towards these regions, which does not allow for a
distinct delineation of the viscous sublayer. This confirms also the
weak nature of the turbulence obtained at this rotational Reynolds
number. This is consistent with previous observations, in particular
from the iso-contours of $Re_t$ and the axial variations of the
magnitude of the two stress vectors $\tau_{tot}$ and $\tau$.

\subsection{Conditional-averaged quadrant analysis}
\begin{figure}[!ht]
\begin{center}
\begin{tabular}{cc}
\scalebox{1}{\includegraphics[width=7.5cm]{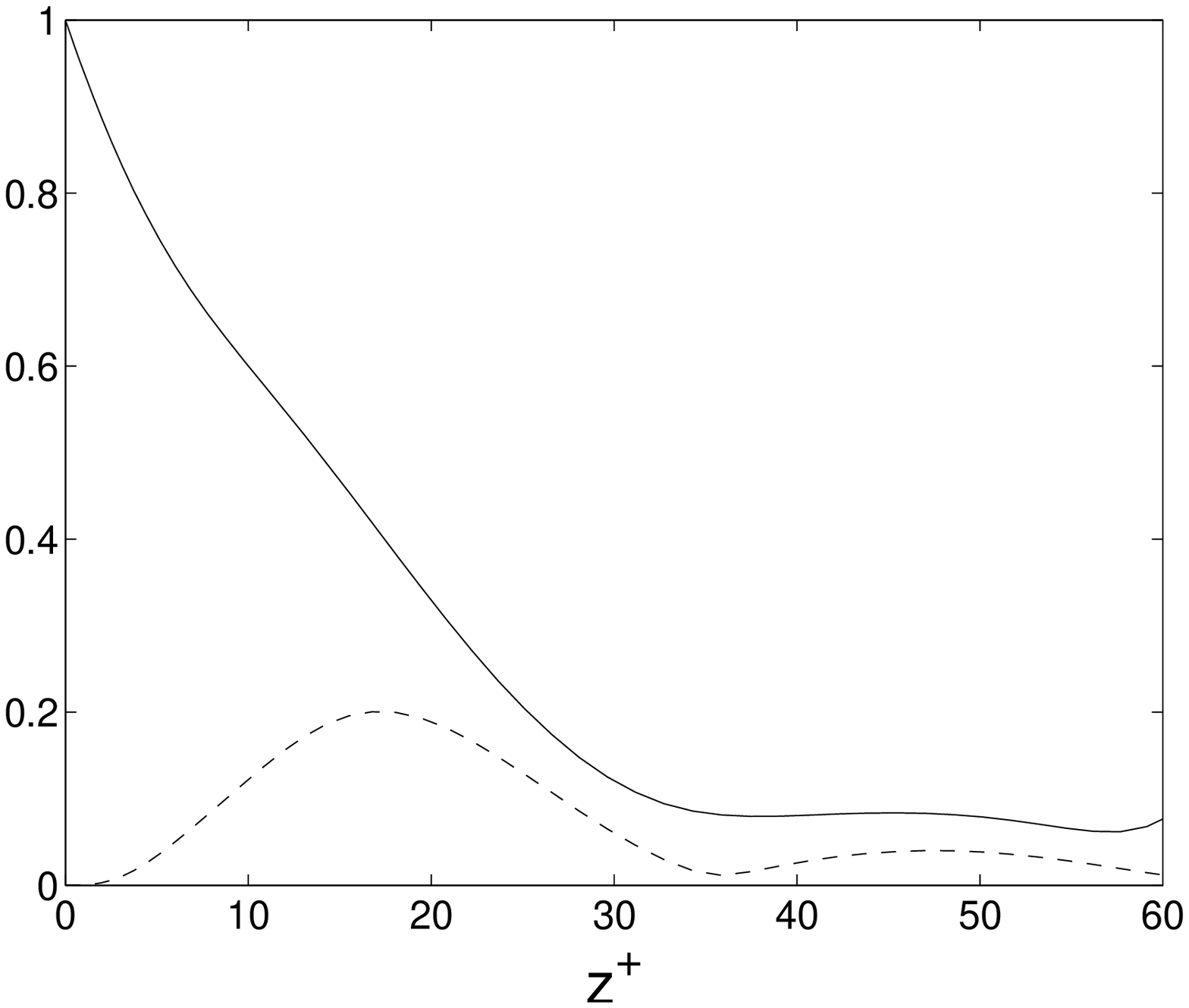}} &
\scalebox{1}{\includegraphics[width=7.5cm]{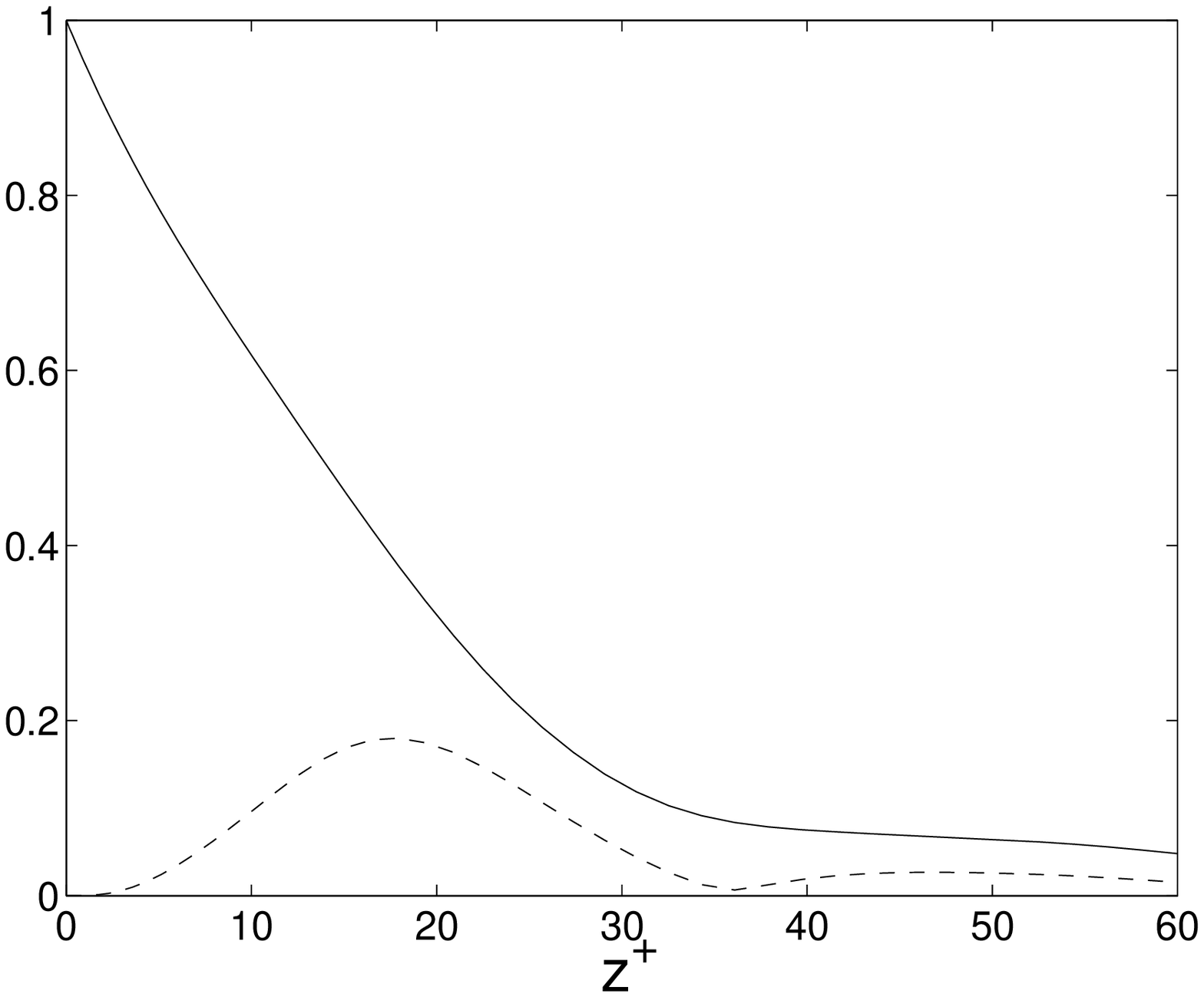}}
\\
(a) & (b)
\end{tabular}
\scalebox{1}{\includegraphics[width=7.5cm]{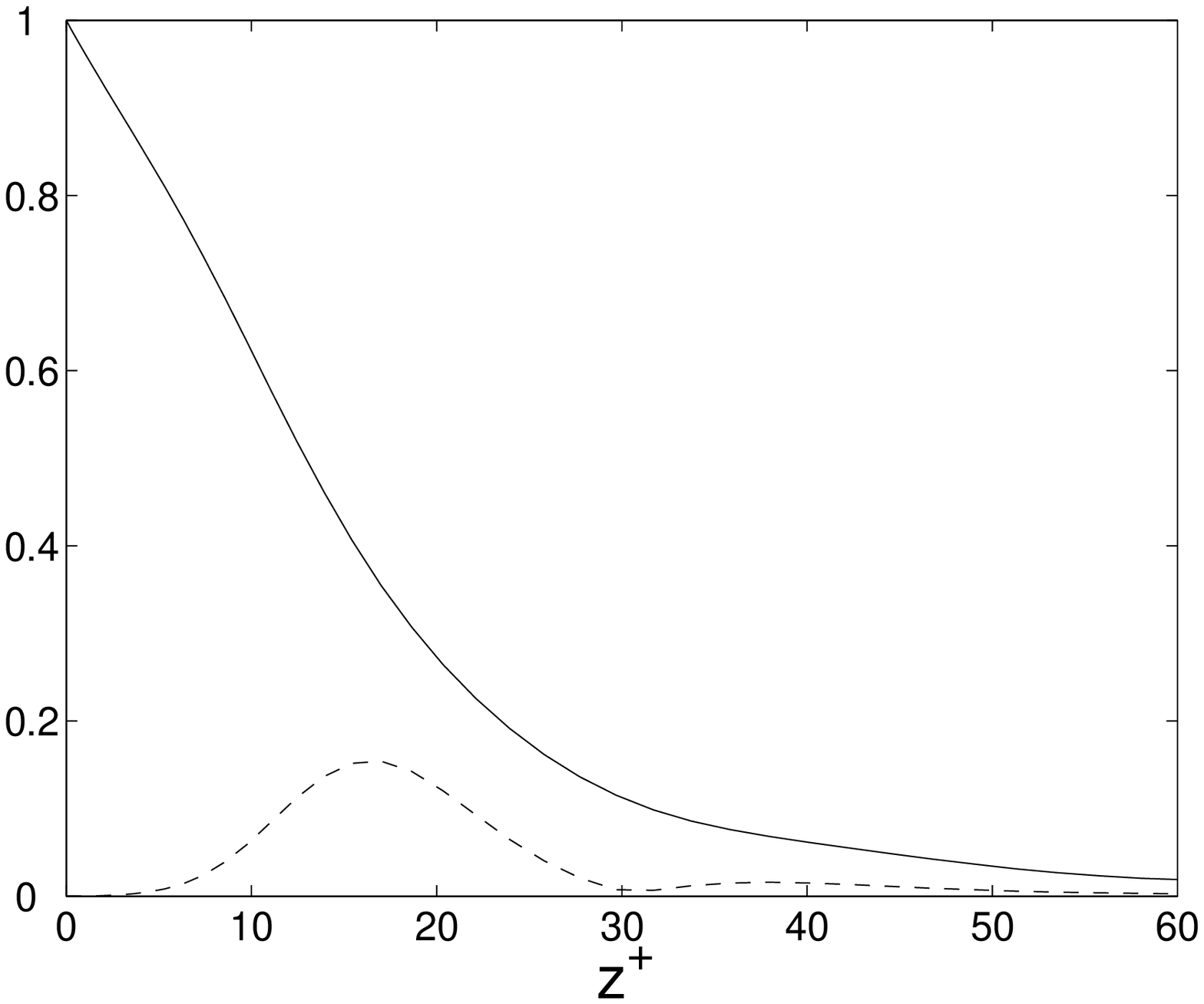}}
\\
(c)
\\
\caption{Randriamampianina et Poncet, Phys. Fluids.} \label{taut}
\end{center}
\end{figure}

\begin{figure}[!ht]
\begin{center}
\scalebox{1}{\includegraphics[width=11.7cm]{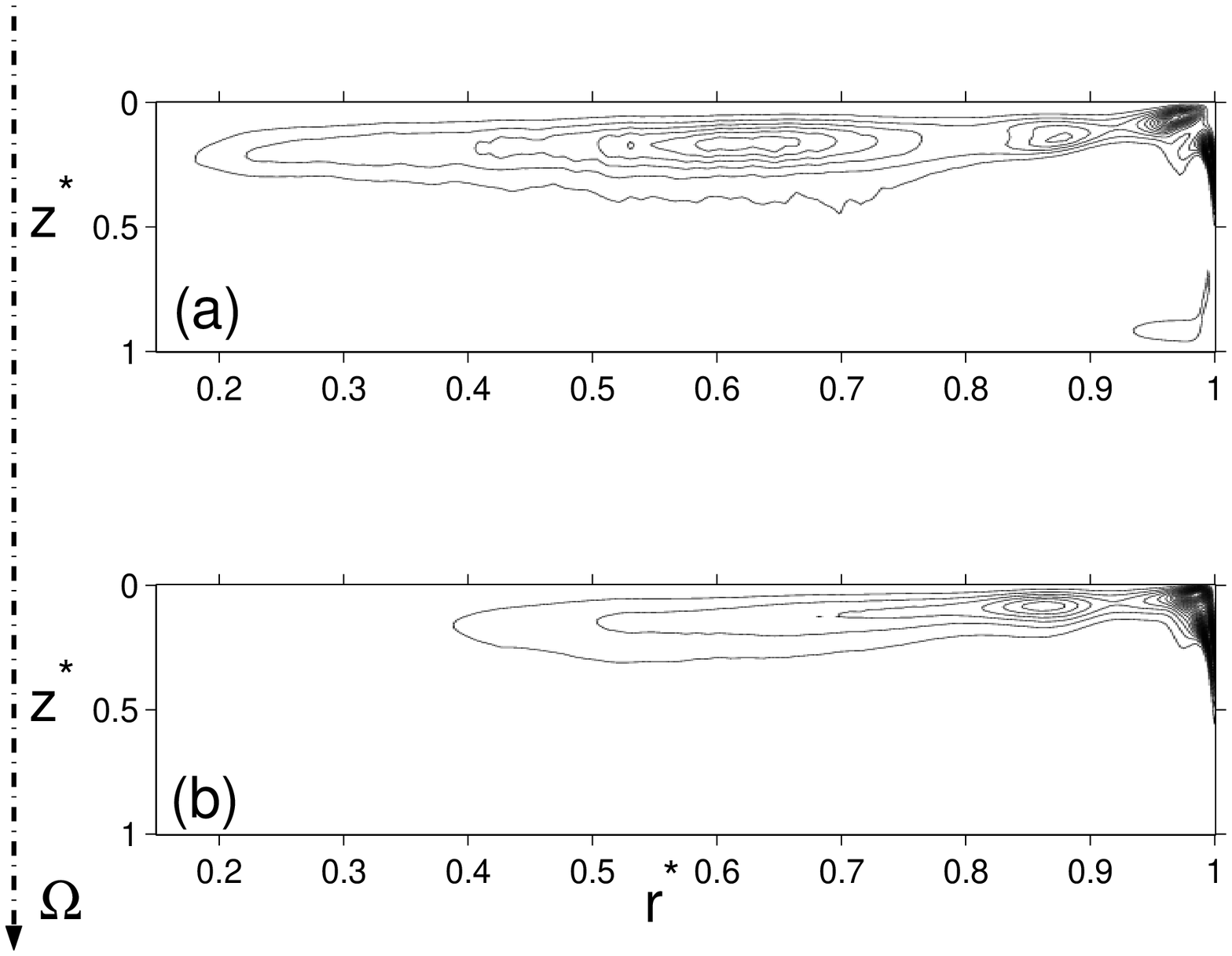}}
\caption{Randriamampianina et Poncet, Phys. Fluids.} \label{kRet}
\end{center}
\end{figure}

\begin{figure}[!ht]
\begin{center}
\vspace{5cm}
\scalebox{1}{\includegraphics[width=8.5cm]{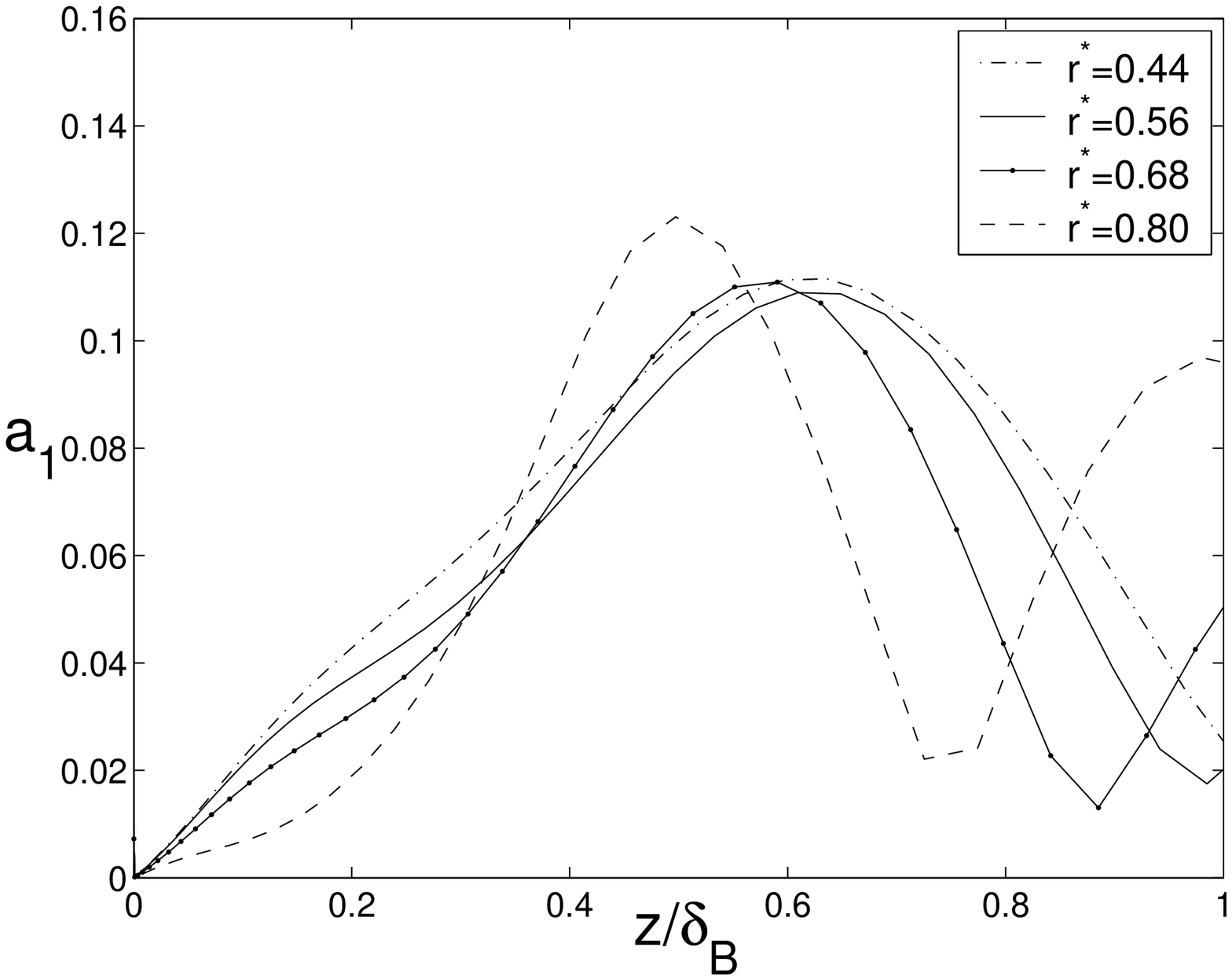}}
\caption{Randriamampianina et Poncet, Phys. Fluids.} \label{town}
\end{center}
\end{figure}

\begin{figure}[!ht]
\centering \vspace{3cm}
\begin{picture}(280,180)

\put(0,0){\includegraphics[width=8cm]{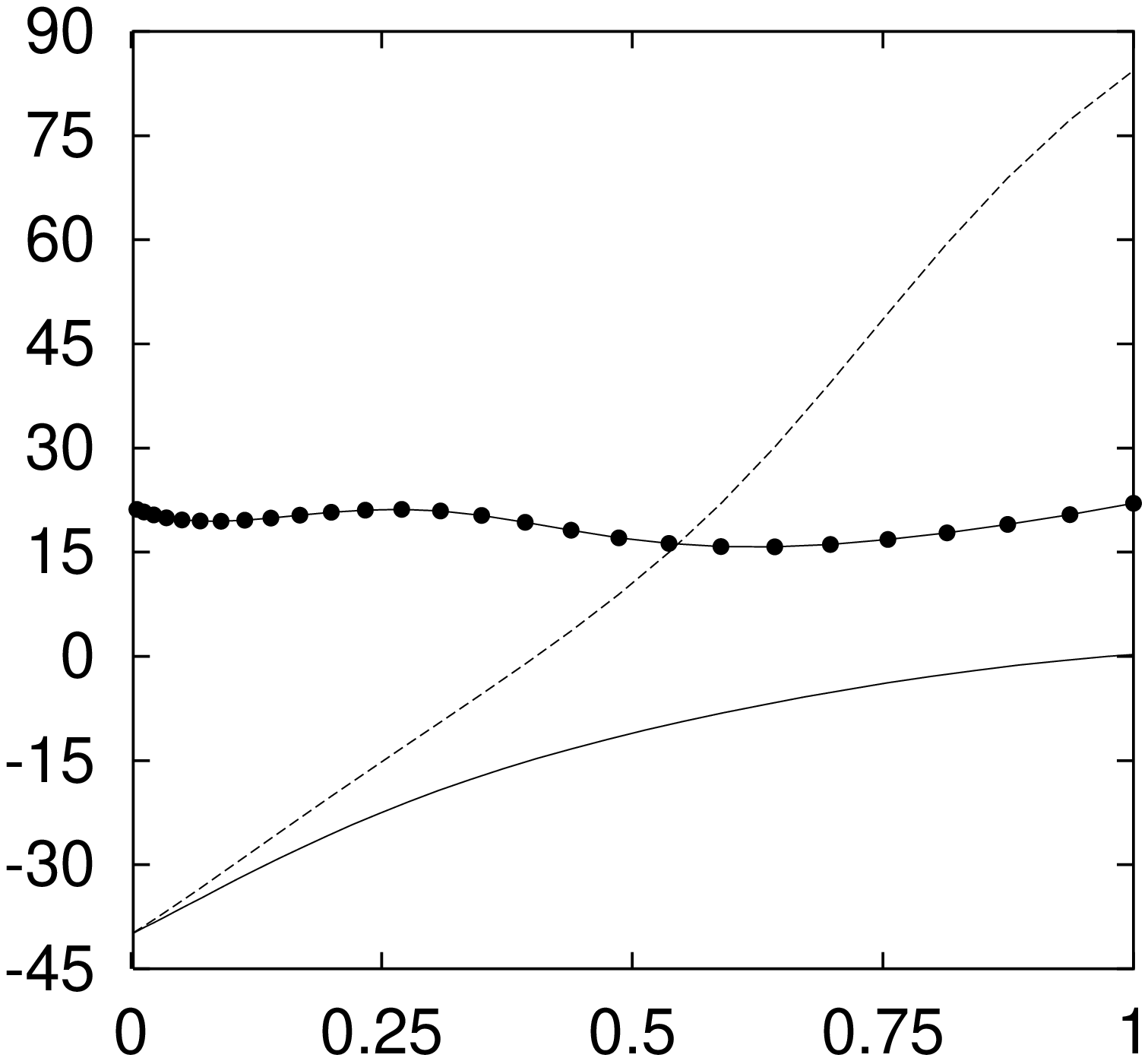}}
\put(120,-15){\shortstack{\large $z/\delta_B$}}
\put(-60,95){\shortstack{\large $Degrees$}}
\end{picture}

\caption{Randriamampianina et Poncet, Phys. Fluids.} \label{gamm}
\end{figure}

\indent Different experimental studies of the flow field near the
wall in a turbulent boundary layer have revealed the occurrence of
intense intermittent bursting events. These have been detected
within the sublayer, and are found to be associated with maximum
levels of the production of turbulence kinetic energy. To gain a
better insight on the near-wall structure of the turbulent boundary
layer along the stator side, a conditional-averaged quadrant
analysis is performed. This provides detailed information on the
Reynolds shear stress producing vortical structures. It corresponds
to four subdivisions of the fluctuations field according to the
combination of the tangential velocity $v'_{\theta}$ and the axial
velocity $v'_z$ \cite{litte94,Kang98}. Following the definitions
given in \cite{lygr01} in a fixed frame, a strong sweep is
associated with $-v'_{\theta}v'_z>\beta
\sqrt{\overline{v'^{2}_{\theta}}} \sqrt{\overline{v'^{2}_{z}}}$ and
$v'_z<0$ (quadrant Q4), and a strong ejection with
$-v'_{\theta}v'_z> \beta \sqrt{\overline{v'^{2}_{\theta}}}
\sqrt{\overline{v'^{2}_{z}}}$ and $v'_z>0$ (quadrant Q2). In the
first quadrant Q1, $v'_{\theta}>0$ and $v'_z>0$, while in the third
quadrant Q3, $v'_{\theta}<0$ and $v'_z<0$. As stated by Littell and
Eaton \cite{litte94}, {\it `an ejection is defined as wall fluid
moving outward and a sweep as outer-layer fluid moving down'} (see
also Robinson \cite{ROB91}). Different criterion levels $\beta$ have
been used in the literature for the conditions imposed to detect
strong ejection and strong sweep \cite{Kang98,lygr01}. Unlike the
geometry considered by Lygren and Andersson \cite{lygr01}, the main
difficulty in the present configuration stems from the confinement,
which yields a dependence with the radial location of all turbulence
quantities, in particular of the wall coordinate $z^+$, as well as
the boundary layer thicknesses. However, we have chosen to fix it in
the present quadrant analysis at the value $z^+=17$ corresponding to
the location of the maximum value of the turbulent shear stress as
seen in figure \ref{taut}. Note that this value of $z^+$ is close to
the one $z^+=20$ used by Lygren and Andersson \cite{lygr01}.
\\
\indent In their experimental study of a fully developed channel
flow, Wallace et $al.$ \cite{wall72} identified the different events
occurring in the wall region from their simultaneous recording of
streamwise and normal velocity components with their product. We
present in figure \ref{ztim80} the space-time maps along the axial
direction of the fluctuating parts of the tangential $v'_{\theta}$
and axial $v'_z$ velocity components and their product
$-v'_{\theta}v'_z$ at the radial location $r^*=0.8$. The figure
shows alternation between positive and negative fluctuations but
with very different strengthes and sizes. The maximum of
fluctuations of the streamwise velocity $v'_{\theta}$ reaches $8\%$
of the mean velocity for the positive part and $10\%$ for the
negative part at this radial location. As expected, the maximum of
the shear stress occurs at about $z^+=17$ (corresponding to
$z^*=0.15$ at $r^*=0.8$), where the magnitude of the shear stress
vector reaches a maximum as observed in figure \ref{taut}.
\\
\indent We display in figures \ref{percent}a and \ref{percent}b the
percentages of respectively strong ejections (Q2) and strong sweeps
(Q4) obtained for three different condition levels in function of
the radial locations $0.56 \le r^* \le 0.85$. Note that levels 1 to
3 corresponds respectively to $\beta$ equal to 1 to 3. These
samplings have been taken independently of the following quadrant
analysis, but are just used to emphasize the levels of the different
quadrants on the total production of turbulence. As reported by Kang
et $al.$ \cite{Kang98}, the percentage of strong ejection events is
higher than that of strong sweep events, and decreases for
increasing values of the criterion level $\beta$. With $\beta=1$,
the percentage of strong ejection events reaches between $1$ to
$7\%$ of the total events for increasing values of $r^*$, while for
the sweep these values fall between $0.5$ to $2\%$. With $\beta=3$,
the percentage for strong ejections event is about 1$\%$, while it
corresponds to less than 0.1$\%$ for strong sweeps, to be compared
with the values obtained by Kang et $al.$ \cite{Kang98} of $2.4$ and
$0.4$ respectively, for the fully turbulent flow over a free
rotating disk. These values confirm the weakness of turbulence
obtained in the present study. Figure \ref{percent}c shows the
percentages of the events in the quadrants Q1 and Q3 at condition
level 2 ($\beta=2$). The most relevant contributions are from the
quadrant Q1 (more than $50\%$ of total events), which contains
motions formed by ejections of high-speed fluids away from the wall,
and the quadrant Q3, which contains inward rushes associated with
sweeps of low-speed fluids. The contributions of the two other
quadrants Q2 and Q4 are very weak (fig.\ref{percent}a and
\ref{percent}b). We have verified that the contributions of the four
quadrants sum up to $1$. According to this analysis, we have
considered the value $\beta=2$ to determine strong events, as used
in \cite{litte94,Kang98,LE00}. Littell and Eaton \cite{litte94} have
also mentioned other conditions, called 'rising' and 'sinking',
based on the sign of the streamwise velocity $v'_{\theta}$, $\mid
v'_{\theta} \mid \le \sqrt{2 \overline{v'^2_{\theta}}}$. However,
these events do not directly act on the shear stress producing
vortical structures.
\\
\indent We display in figures \ref{vwevent} the variation with
$-\Delta r^+$ ($r^+ = r v_{\tau}/\nu$) of the conditionally averaged
Reynolds shear stress normalized by the unconditionally ensemble
averaged Reynolds shear stress $<v'_{\theta}v'_z>$ near a strong
ejection (fig.\ref{vwevent}a) or a strong sweep
(fig.\ref{vwevent}b). The contributions of each quadrant are also
presented. The choice of the separation distance $-\Delta r^+$ in
the radial direction is motivated by the works of Lygren and
Andersson \cite{lygr01} to allow direct comparisons of the two
results. On the other hand, for the range of radial locations where
the analysis is performed, $0.56 \le r^* \le 0.8$, the angle of the
mean velocity $\gamma_m=tan^{-1}(V_r/V_{\theta})$ remains small:
$\gamma_m \sim -18^{\circ}$ at the wall distance chosen $z^+=17$
corresponding to $z/\delta_B=0.365$ at $r^*=0.68$ (fig.\ref{gamm}).
Thus very similar behaviors are expected when using the local
spanwise direction.
\\
\indent The profiles in figures \ref{vwevent}-\ref{vwsweep} exhibit
the main features reported in previous related works from
experiments on a rotating free disk \cite{litte94,Kang98} and from
simulations on infinite rotor-stator systems \cite{lygr01,LYAN04}.
The center peak in each plot, concerning a strong sweep or ejection,
is associated with two secondary peaks generated by the opposite
event. Kang et $al.$ \cite{Kang98} reported that these peaks
represent a pair of streamwise vortices generating a strong event.
The center peak contains the combined effect of both vortices, while
the secondary peaks contain the effect of one single vortex. Thus
the asymmetries observed by Littell and Eaton \cite{litte94} or
Lygren and Andersson \cite{lygr01} can be discerned by comparing the
secondary peaks \cite{LE00}.
\\
\indent Beyond the confinement of the geometry, another specific
characteristics of the present study comes from the weakness of the
turbulence obtained. This leads to differences on levels compared
with the cited references. However, it is worth to mention that
individual contributions to the shear stress as large as $50
\hspace{0.2cm} \overline{v'_{\theta} v'_z}$ have been identified
during the present simulation. As observed by \cite{Kang98} from
their experimental study on a free rotating disk, the conditionally
shear stress matches the unconditionally shear stress at large
values of $\mid \Delta r^+ \mid$, giving a ratio 1.0 in the figures,
as the conditionally shear stress at these spanwise distances
becomes independent of the event produced at $\Delta r^+=0$. Strong
ejection and sweep are directly associated with the near-wall
vortical motion. An ejection event (at $\Delta r^+=0$) produced by a
near-wall streamwise vortex is associated with two sweeps located
symmetrically at $\Delta r^+ \simeq \pm 20$. Such a combination can
be seen from the space-time maps of the instantaneous shear stress
in figure \ref{ztim80}, which show alternation between positive and
negative parts of different strengthes. Kang et $al.$ \cite{Kang98}
concluded that clockwise and counter-clockwise vortices presented
the same characteristics according to the behaviors of conditionally
averaged streamwise and wall-normal velocities, in contrast with the
conclusions of Littell and Eaton \cite{litte94} and Lygren and
Andersson \cite{lygr01}. In the present case, the asymmetries on
strengthes of neighbouring events are observed similarly to the
findings of Lygren and Andersson \cite{lygr01}, who concluded that
clockwise vortices contribute much more to the Reynolds shear stress
than counter-clockwise vortices. The same behavior applies in the
presence of a sweep event. In this case, the levels of the
surrounding ejections approach the strong sweep level and are even
slightly beyond the fixed criterion condition $\beta=2$, as seen in
figure \ref{vwevent}b, while the levels of sweeps around a strong
ejection are less important (fig.\ref{vwevent}a), in agreement with
the results reported by Lygren and Andersson \cite{lygr01} from
simulation of fully turbulent rotor-stator flows. In particular,
these authors also obtained a level of an ejection close to the
condition criterion $\beta=2$ in the case of a strong sweep event.
These give an indication on the size and the strength of the
vortical structures in the vicinity of a strong event. The figures
clearly show that the ejection (Q2) and sweep (Q4) quadrants
contribute much more to the Reynolds shear stress production than
the two other quadrants: the total averaged profiles follow
practically the same profiles as the contributions from quadrants Q2
and Q4 during the different events. On the other hand, it seems that
the weakness of the turbulence in the present simulation accentuates
the features observed in previous works.
\\
\indent To fix the effects of the contributions of each quadrant on
the presence of these asymmetries in the vicinity of a strong event,
we present in figures \ref{vweject}-\ref{vwsweep} the variation with
$-\Delta r^+$ of the conditionally averaged streamwise and
wall-normal velocity components. These have been normalized by the
corresponding root-mean-square of the unconditioned velocity
fluctuations. As expected the extrema occur at $ \Delta r^+ \simeq
\pm 20$. Asymmetries are observed from the contributions of each
quadrant, but the contributions from quadrants Q2 and Q4,
responsible for generating the ejection and sweep events, are
clearly larger than the two others, which have less relation to the
streamwise vortical structures. Even though in the vicinity of a
strong sweep (fig.\ref{vwsweep}b) the quadrant Q1 contributes in
addition to quadrant Q2 to give two nearly symmetric ejections of
same strength for the wall-normal velocity component, the
corresponding streamwise velocity profile shows that asymmetries
mainly result from the ejection quadrant Q2. These behaviors are
reflected in the conditionally averaged shear stresses, where the
contributions of quadrants Q1 and Q3 are significantly less
important on the total averaged components. Therefore, the present
results support the conclusions proposed by Lygren and Andersson
\cite{lygr01}: Case 1 vortices generate more special events than
Case 2 vortices. Although the B\"odewadt boundary layer is known to
possess its own charateristics compared with the Ekman boundary
layer studied by Kang et $al.$ \cite{Kang98}, the present behaviors
can not be only attributed to such differences. It is worth to
recall the weakness of the turbulence obtained with the rotation
rate considered, and confinement may also play a role in this
analysis, with variation of turbulence quantities with radius.

\section{Conclusions}
Experimental investigations have been performed and compared to DNS
calculations to describe the turbulent flow in an enclosed
rotor-stator cavity of very large aspect ratio $G=18.32$. The
rotational Reynolds number under consideration in the present work
is fixed to $Re=9.5 \times 10^4$.
\\
\indent The flow belongs to the Batchelor family. It is divided into
three distinct zones: two boundary layers separated by a central
rotating core. The entrainment coefficient $K$ of the fluid ranges
from $0.375$ to $0.418$ close to the theoretical value 0.431
\cite{owen89} and the empirical value of $0.438$ for fully turbulent
flows proposed by Poncet et $al.$ \cite{Pon04}. The polar profile
falls between the typical fully turbulent flow \cite{lygr01} and the
laminar solution of Von K\'arm\'an \cite{VKT21} and indicates that
the Ekman layer is laminar, whereas the B\"odewadt layer is
turbulent. The computed results are found here in excellent
agreement with the velocity measurements for the mean field.
\\
\indent The study includes also turbulence measurements, which were
seldom possible in previous works of the literature. It appears that
the turbulence intensities $R_{rr}^*$ and $R_{\theta\theta}^*$ in
the B\"odewadt layer decrease from the periphery to the center of
the cavity, whereas the Ekman layer remains laminar. The
$R_{r\theta}^*$ component is close to zero, indicating that there is
practically no turbulent shear stress in the whole cavity. Although
the profiles from the simulation resemble the behavior obtained from
measurements, a slight discrepancy observed towards the periphery
results from the different closures at the junction between the
rotating disk and the stationary outer casing, where turbulence
prevails according to the isocontours of the turbulent Reynolds
number. An approximately self-similar behavior is obtained in the
B\"odewadt layer for $0.56 \le r^* \le 0.8$. The reduction of the
Townsend structural parameter $a_1$ below the limiting value $0.15$
and the variation in the wall-normal direction of the different
characteristic angles confirm the three-dimensional turbulent nature
of the flow along the stator wall.
\\
\indent The turbulence kinetic energy budgets reveal that production
is the major contribution with a maximum obtained for $z^+ \simeq
12$ independently of the radial location, confirming the
self-similar behavior of the B\"odewadt layer. Towards the outer
stationary casing, an increasing level of the viscous diffusion is
observed, in complement of the dissipation, to balance the
production, which shows the weak level of turbulence obtained at the
rotation rate considered.
\\
\indent Finally, a quadrant analysis is performed. The asymmetries
observed by different authors in 3DTBLs with rotation
(\cite{litte94},\cite{Kang98},\cite{LE00},\cite{lygr01}) have been
clearly detected and the analysis of conditionally averaged
streamwise and wall-normal velocity components confirms that these
asymmetries mainly arise from the contributions of quadrants Q2 and
Q4, responsible for the generation of ejection and sweep events.
Moreover, Case 1 vortices are found to be the major source of
generation of special strong events, in the present study
characterized by a weak turbulence level and confinement. This
result is in agreement with the conclusions of Lygren and Andersson
\cite{lygr01} in an "infinite" rotor-stator system, unlike the case
reported for three-dimensional turbulent Ekman boundary layers
\cite{litte94,LE00}.

\begin{acknowledgements}
Numerical computations have been carried out on the NEC SX-5 (IDRIS,
Orsay, France). Financial supports for the experimental approach
from SNECMA Moteurs, Large Liquid Propulsion (Vernon, France) are
also gratefully acknowledged. The authors thank Dr. Roland Schiestel
and Dr. Marie-Pierre Chauve (IRPHE, Marseille, France) for fruitful
discussions. They appreciate also the numerous valuable suggestions
and comments on the manuscript provided by the reviewers.
\end{acknowledgements}

\bibliographystyle{unsrt}

\begin{thebibliography}{10}

\bibitem{lygr01}
M.~Lygren and H.~I. Andersson.
\newblock Turbulent flow between a rotating and a stationary disk.
\newblock {\em J. Fluid. Mech.}, 426:297--326, 2001.

\bibitem{Kang98}
H.~S. Kang, H.~Choi, and J.~Y. Yoo.
\newblock On the modification of the near-wall coherent structure in a
  three-dimensional turbulent boundary layer on a free rotating disk.
\newblock {\em Phys. Fluids}, 10(9):2315--2322, 1998.

\bibitem{batc51}
G.~K. Batchelor.
\newblock Note on a class of solutions of the \mbox{Navier-Stokes} equations
  representing steady rotationally-symmetric flow.
\newblock {\em Q. J. Mech. Appl. Math.}, 4:29--41, 1951.

\bibitem{Stew53}
K.~Stewartson.
\newblock On the flow between two rotating coaxial disks.
\newblock {\em Proc. Camb. Phil. Soc.}, 49:333--341, 1953.

\bibitem{Mel68}
G.L. Mellor, P.J. Chapple, and V.K. Stokes.
\newblock On the flow between a rotating and a stationary disk.
\newblock {\em J. Fluid. Mech.}, 31(1):95--112, 1968.

\bibitem{dail60}
J.~W. Daily and R.~E. Nece.
\newblock Chamber dimension effects on induced flow and frictional resistance
  of enclosed rotating disks.
\newblock {\em ASME J. Basic Eng.}, 82:217--232, 1960.

\bibitem{owen89}
J.~M. Owen and R.~H. Rogers.
\newblock {\em Flow and Heat Transfer in Rotating-Disc Systems - Vol.1:
  Rotor-Stator Systems}.
\newblock Ed. Morris, W.D. John Wiley and Sons Inc., New-York, 1989.

\bibitem{itoh92}
M.~Itoh, Y.~Yamada, S.~Imao, and M.~Gonda.
\newblock Experiments on turbulent flow due to an enclosed rotating disk.
\newblock {\em Exp. Thermal Fluid Sci.}, 5:359--368, 1992.

\bibitem{chea94}
S.C. Cheah, H.~Iacovides, D.C. Jackson, H.~Ji, and B.E. Launder.
\newblock Experimental investigation of enclosed rotor-stator disk flows.
\newblock {\em Exp. Therm. Fluid Sci.}, 9:445--455, 1994.

\bibitem{WILK85}
S.P. Wilkinson and M.R. Malik.
\newblock Stability experiments in the flow over a rotating disk.
\newblock {\em AIAA J.}, 23(4):588--595, 1985.

\bibitem{koba94}
R.~Kobayashi.
\newblock Review: laminar-turbulent transition of three-dimensional boundary
  layers on rotating bodies.
\newblock {\em J. Fluid Eng.}, 116:200--211, 1994.

\bibitem{gaut99}
G.~Gauthier, P.~Gondret, and M.~Rabaud.
\newblock Axisymmetric propagating vortices in the flow between a stationary
  and a rotating disk enclosed by a cylinder.
\newblock {\em J. Fluid Mech.}, 386:105--126, 1999.

\bibitem{scho01}
L.~Schouveiler, P.~Le~Gal, and M.-P. Chauve.
\newblock Instabilities of the flow between a rotating and a stationary disk.
\newblock {\em J. Fluid Mech.}, 443:329--350, 2001.

\bibitem{IYIG90}
M.~Itoh, Y.~Yamada, S.~Imao, and M.~Gonda.
\newblock Experiments on turbulent flow due to an enclosed rotating disk.
\newblock In W.~Rodi and E.N. Ganic, editors, {\em Engineering Turbulence
  Modeling and Experiments}, pages 659--668, New-York, 1990. Elsevier.

\bibitem{Pon04}
S.~Poncet, M.P. Chauve, and P.~Le~Gal.
\newblock Turbulent rotating disk with inward throughflow.
\newblock {\em J. Fluid. Mech.}, 522:253--262, 2005.

\bibitem{PON05}
S.~Poncet, M.~P. Chauve, and R.~Schiestel.
\newblock Batchelor versus stewartson flow structures in a rotor-stator cavity
  with throughflow.
\newblock {\em Phys. Fluids}, 17(7), 2005.

\bibitem{ES96}
L.~Elena and R.~Schiestel.
\newblock Turbulence modeling of rotating confined flows.
\newblock {\em Int. J. Heat Fluid Flow}, 17:283--289, 1996.

\bibitem{serr01}
E.~Serre, E.~Crespo~del Arco, and P.~Bontoux.
\newblock Annular and spiral patterns in flows between rotating and stationary
  discs.
\newblock {\em J. Fluid. Mech.}, 434:65--100, 2001.

\bibitem{JAC02}
R.~Jacques, P.~Le~Qu\'er\'e, and O.~Daube.
\newblock Axisymmetric numerical simulations of turbulent flow in a rotor
  stator enclosures.
\newblock {\em Int. J. Heat Fluid Flow}, 23:381--397, 2002.

\bibitem{serr04}
E.~Serre, E.~Tuliszka-Sznitko, and P.~Bontoux.
\newblock Coupled numerical and theoretical study of the flow transition
  between a rotating and a stationary disk.
\newblock {\em Phys. Fluids}, 16(3):688--706, 2004.

\bibitem{LYAN04}
M.~Lygren and H.~I. Andersson.
\newblock Large eddy simulations of the turbulent flow between a rotating and a
  stationary disk.
\newblock {\em Z. Angew. Math. Phys.}, 55:268--281, 2004.

\bibitem{PONTHE}
S.~Poncet.
\newblock {\em \'Ecoulements de type rotor-stator soumis \`a un flux axial: de
  Batchelor \`a Stewartson}.
\newblock PhD thesis, Universit\'e Aix-Marseille I, 2005.

\bibitem{JOHN96}
J.P. Johnston and K.A. Flack.
\newblock Review - advances in three-dimensional turbulent boundary layers with
  emphasis on the wall-layer regions.
\newblock {\em J. Fluids Engng}, 118:219--232, 1996.

\bibitem{saric03}
W.S. Saric, H.L. Reed, and E.B. White.
\newblock Stability and transition of three-dimensional boundary layers.
\newblock {\em Ann. Rev. Fluid Mech.}, 35:413--440, 2003.

\bibitem{ROB91}
S.K. Robinson.
\newblock Coherent motions in the turbulent boundary layer.
\newblock {\em Ann. Rev. Fluid Mech.}, 23:601--639, 1991.

\bibitem{WEBS96}
D.~Webster, D.~Degraaff, and J.K. Eaton.
\newblock Turbulence characteristics of a boundary layer over a swept bump.
\newblock {\em J. Fluid. Mech.}, 323:1--22, 1996.

\bibitem{cori69}
E.~R. Corino and R.~S. Brodkey.
\newblock A visual investigation of the wall region in turbulent flow.
\newblock {\em J. Fluid Mech.}, 37(1):1--30, 1969.

\bibitem{kim71}
H.~T. Kim, S.~J. Kline, and W.~C. Reynolds.
\newblock The production of turbulence near a smooth wall in a turbulent
  boundary layer.
\newblock {\em J. Fluid Mech.}, 50(1):133--160, 1971.

\bibitem{EAT95}
J.K. Eaton.
\newblock Effects of mean flow three-dimensionality on turbulent boundary-layer
  structure.
\newblock {\em AIAA J.}, 33:2020--2025, 1995.

\bibitem{SHI92}
T.~Shizawa and J.K. Eaton.
\newblock Turbulence measurements for a longitudinal vortex interacting with a
  three-dimensional turbulent boundary layer.
\newblock {\em AIAA J.}, 30:49--55, 1992.

\bibitem{litte94}
H.~S. Littell and J.~K. Eaton.
\newblock Turbulence characteristics of the boundary layer on a rotating disk.
\newblock {\em J. Fluid. Mech.}, 266:175--207, 1994.

\bibitem{wu2000}
X.~Wu and K.~D. Squires.
\newblock Prediction and investigation of the turbulent flow over a rotating
  disk.
\newblock {\em J. Fluid. Mech.}, 418:231--264, 2000.

\bibitem{CHIA96}
C.~Chiang and J.K. Eaton.
\newblock An experimental study of the effects of three-dimensionality on the
  near wall turbulence structures using flow visualization.
\newblock {\em Exps Fluids}, 20:266--272, 1996.

\bibitem{FLAC97}
K.A. Flack.
\newblock Near-wall structure of three-dimensional turbulent boundary layers.
\newblock {\em Exps Fluids}, 23:335--340, 1997.

\bibitem{LE00}
A.T. Le, G.N. Coleman, and J.~Kim.
\newblock Near-wall turbulence structures in three-dimensional boundary layers.
\newblock {\em Int. J. Heat Fluid Flow}, 21:480--488, 2000.

\bibitem{BAT76}
C.J. Bates and T.D. Hughes.
\newblock Real-time statistical ldv system for the study of a high reynolds
  number, low turbulence intensity flow.
\newblock {\em J. Phys. E. Sci. Instrum.}, 9:955--958, 1976.

\bibitem{zang90}
T.~A. Zang.
\newblock Spectral methods for simulations of transition and turbulence.
\newblock {\em Comp. Meth. Appl. Mech. Eng.}, 80:209--221, 1990.

\bibitem{tave91}
S.~J. Tavener, T.~Mullin, and K.~A. Cliffe.
\newblock Novel bifurcation phenomena in a rotating annulus.
\newblock {\em J. Fluid. Mech.}, 229:483--497, 1991.

\bibitem{rand97}
A.~Randriamampianina, L.~Elena, J.~P. Fontaine, and R.~Schiestel.
\newblock Numerical prediction of laminar, transitional and turbulent flows in
  shrouded rotor-stator systems.
\newblock {\em Phys. Fluids}, 9(6):1696--1713, 1997.

\bibitem{rand01}
A.~Randriamampianina, R.~Schiestel, and M.~Wilson.
\newblock Spatio-temporal behaviour in an enclosed corotating disk pair.
\newblock {\em J. Fluid Mech.}, 434:39--64, 2001.

\bibitem{canu87}
C.~Canuto, M.~Y. Hussaini, A.~Quarteroni, and T.~A. Zang.
\newblock {\em Spectral methods in fluid dynamics}.
\newblock Springer Verlag, Berlin, 1987.

\bibitem{vane86}
J.~M. Vanel, R.~Peyret, and P.~Bontoux.
\newblock A pseudospectral solution of vorticity-stream function equations
  using the influence matrix technique.
\newblock In K.W. Morton and M.J. Baines, editors, {\em Num. Meth. Fluid
  Dynamics II}, pages 463--475, Clarendon, 1986.

\bibitem{hugth}
S.~Hugues.
\newblock {\em D\'eveloppement d'un algorithme de projection pour m\'ethodes
  pseudospectrales: application \`a la simulation d'instabilit\'es
  tridimensionnelles dans les cavit\'es tournantes. Mod\'elisation
  d'\'ecoulements turbulents dans les syst\`emes rotor-stator.}
\newblock PhD thesis, Universit\'e Aix-Marseille II, 1998.

\bibitem{rasp02}
I.~Raspo, S.~Hugues, E.~Serre, A.~Randriamampianina, and P.~Bontoux.
\newblock A spectral projection method for the simulation of complex
  three-dimensional rotating flows.
\newblock {\em Computers and Fluids}, 31:745--767, 2002.

\bibitem{hugu98}
S.~Hugues and A.~Randriamampianina.
\newblock An improved projection scheme applied to pseudospectral methods for
  the incompressible \mbox{Navier-Stokes} equations.
\newblock {\em Int. J. Numer. Meth. Fluids}, 28:501--521, 1998.

\bibitem{hald84}
P.~Haldenwang, G.~Labrosse, S.~Abboudi, and M.~Deville.
\newblock Chebyshev 3-d spectral and 2-d pseudospectral solvers for the
  \mbox{Helmholtz} equation.
\newblock {\em J. Comput. Phys.}, 55:115--128, 1984.

\bibitem{VKT21}
T.~Von~K\'arm\'an.
\newblock Uber laminare und turbulente \mbox{Reibung}.
\newblock {\em Z. Angew. Math. Mech.}, 1:233--252, 1921.

\bibitem{Pon05cras}
S.~Poncet and A.~Randriamampianina.
\newblock \'ecoulement turbulent dans une cavit\'e rotor-stator ferm\'ee de
  grand rapport d'aspect.
\newblock {\em C.R. M\'ecanique}, 333:783--788, 2005.

\bibitem{cole2000}
G.~N. Coleman, J.~Kim, and P.R. Spalart.
\newblock A numerical study of strained three-dimensional wall-bounded
  turbulence.
\newblock {\em J. Fluid Mech.}, 416:75--116, 2000.

\bibitem{WILL72}
W.W. Willmarth and S.S. Lu.
\newblock Structure of the reynolds stress near the wall.
\newblock {\em J. Fluid. Mech.}, 55:65--92, 1972.

\bibitem{wall72}
J.~M. Wallace, H.~Eckelmann, and R.~S. Brodkey.
\newblock The wall region in turbulent shear flow.
\newblock {\em J. Fluid. Mech.}, 54:39--48, 1972.

\end{thebibliography}

\clearpage

\begin{itemize}
\item Fig.1: Schematic diagram of the experimental rig with relevant
notations.
\item Fig.2: Axial variations of the mean radial $V_r^*$ and tangential $V_{\theta}^*$
velocity components at four radial locations : (a) $r^*=0.44$, (b)
$r^*=0.56$, (c) $r^*=0.68$, (d) $r^*=0.8$. Comparisons between the
3D simulation (solid lines) and the LDA measurements (symbols).
\item Fig.3: Polar plot of the velocity distribution in the whole gap
between the disks.
\item Fig.4: Computed ($-$) and measured ($\circ$) axial variations of
two normal Reynolds stress tensor components $R_{rr}^*$ and
$R_{\theta\theta}^*$ and one shear Reynolds stress tensor component
$R_{r\theta}^*$ at four radial locations : (a) $r^*=0.44$, (b)
$r^*=0.56$, (c) $r^*=0.68$, (d) $r^*=0.8$.
\item Fig.5: Reynolds stresses near the stator wall at three radial
locations in function of the wall coordinate $z^+$. Normalized with
$v_{\tau}$ at the wall with $v_{\tau}/(\Omega b)=2.14 \times
10^{-2}$ at $r^*=0.56$, $v_{\tau}/(\Omega b)=2.44 \times 10^{-2}$ at
$r^*=0.68$, and $v_{\tau}/(\Omega b)=2.95 \times 10^{-2}$ at
$r^*=0.80$.
\item Fig.6: Magnitude of the turbulent shear stress vector $\tau$
(dashed) and magnitude of the total shear stress vector $\tau_{tot}$
(solid), normalized with $v_{\tau}$ at the wall, at three radial
locations: (a) $r^*=0.56$ with $v^2_{\tau}/(\Omega b)^2=4.58 \times
10^{-4}$, (b) $r^*=0.68$ with $v^2_{\tau}/(\Omega b)^2=5.93 \times
10^{-4}$, (c) $r^*=0.8$ with $v^2_{\tau}/(\Omega b)^2=8.68 \times
10^{-4}$.
\item Fig.7: $23$ regularly spaced isocontours (a) of the turbulent
Reynolds number $0 \leq Re_t=k^2/(\nu \epsilon) \leq 20.52$ and (b)
of the turbulent kinetic energy $0 \leq k^*=k/(\Omega b)^2 \leq
1.377 \times 10^{-2}$ (apparent aspect ratio equal to $5$).
\item Fig.8: Axial variation of the Townsend structural parameter $a_1$
in the stator side boundary layer at four radial locations.
\item Fig.9: Axial variation at $r^*=0.68$ of the mean velocity angle $\gamma_m$
(solid line), the mean gradient angle $\gamma_g$ (dotted line) and
the Reynolds shear stress angle $\gamma_{\tau}$ ($\bullet$).
\item Fig.10: Budgets for the turbulence kinetic energy normalized by $2 \Omega^3 h^2$ at three radial
locations: (a) $r^*=0.56$, (b) $r^*=0.68$, (c) $r^*=0.8$.
\item Fig.11: Space-time maps of the fluctuations of the tangential $v'_{\theta}$ and axial $v'_z$
velocity components and $-v'_{\theta}v'_z$ at $r^*=0.8$: positive
(solid) and negative (dashed) values.
\item Fig.12: Percentages in function of the radial locations of the strong events at different
condition levels: (a) strong ejections (Q2), (b) strong sweeps (Q4)
and (c) percentages of the events in quadrants Q1 and Q3 at
condition level 2 ($\beta=2$).
\item Fig.13: Conditionally averaged Reynolds shear stress at $z^+=17$ in the vicinity of a
strong ejection event $(a)$ and of a strong sweep event $(b)$.
$\bullet$ sum of all quadrant quantitities.
\item Fig.14: Conditionally averaged velocity components at $z^+=17$ in the vicinity of a strong
ejection event $(a)$ streamwise velocity and $(b)$ wall-normal
velocity. $\bullet$ sum of all quadrant quantitities.
\item Fig.15: Conditionally averaged velocity components at $z^+=17$ in the vicinity of a strong
sweep event $(a)$ streamwise velocity and $(b)$ wall-normal
velocity. $\bullet$ sum of all quadrant quantitities.
\end{itemize}

\clearpage

\newpage

\newpage

\newpage
\newpage

\newpage

\newpage

\newpage

\newpage

\newpage
\begin{figure}[!ht]
\centering
\vspace{3cm}
\begin{tabular}{cc}
\includegraphics[width=7.5cm]{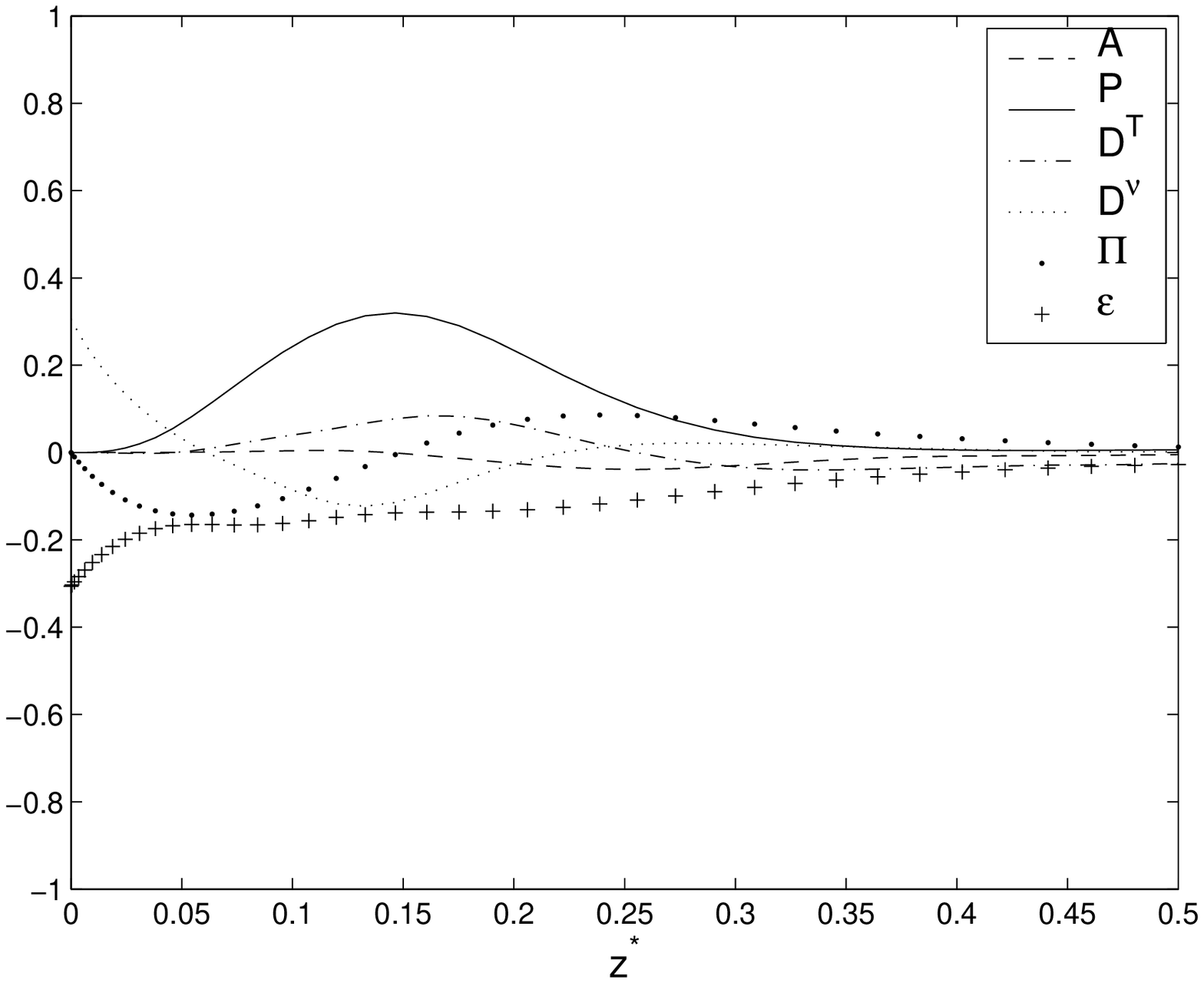} &
\includegraphics[width=7.5cm]{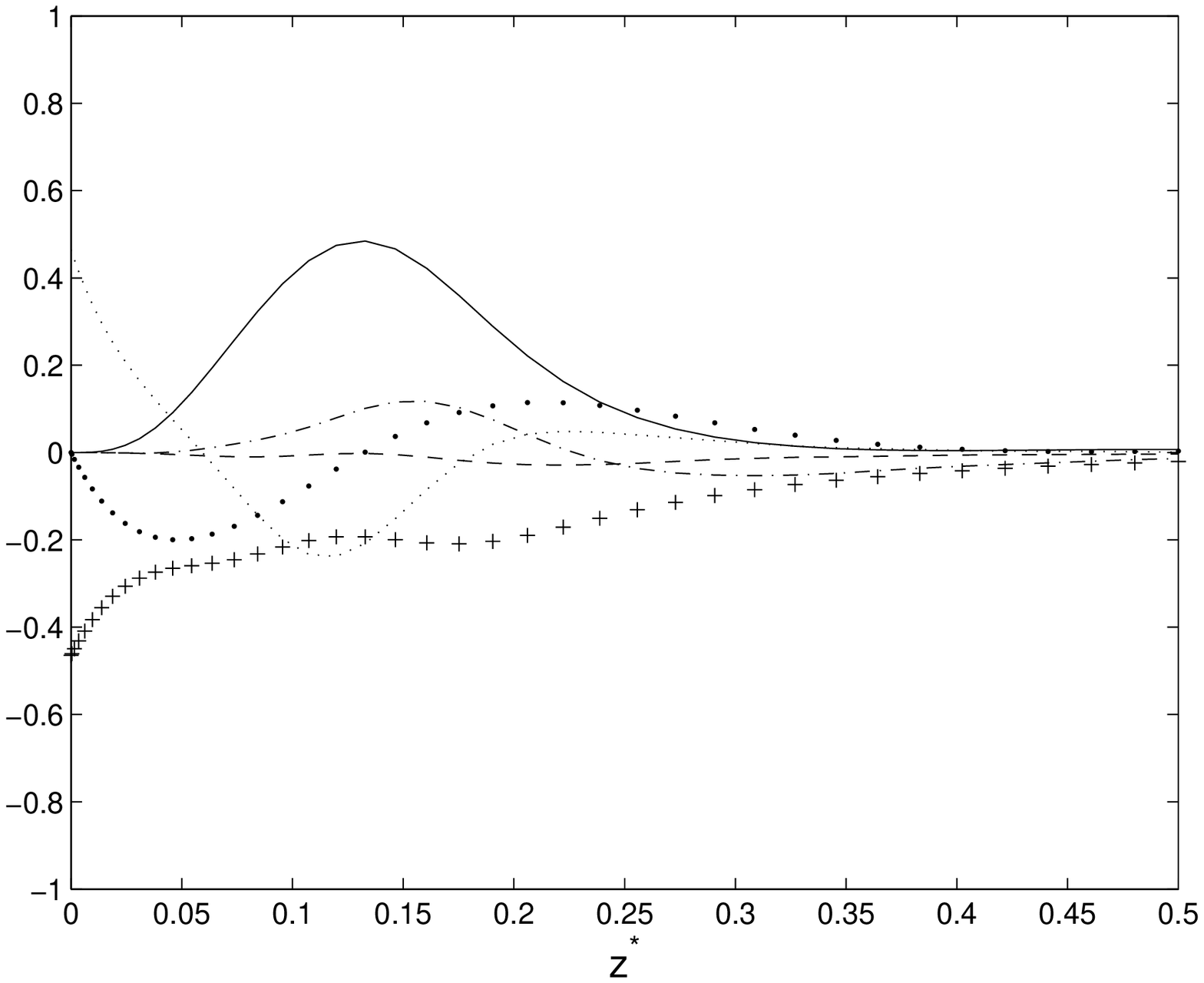}
\\
(a) & (b)
\end{tabular}
\begin{tabular}{c}
\includegraphics[width=7.5cm]{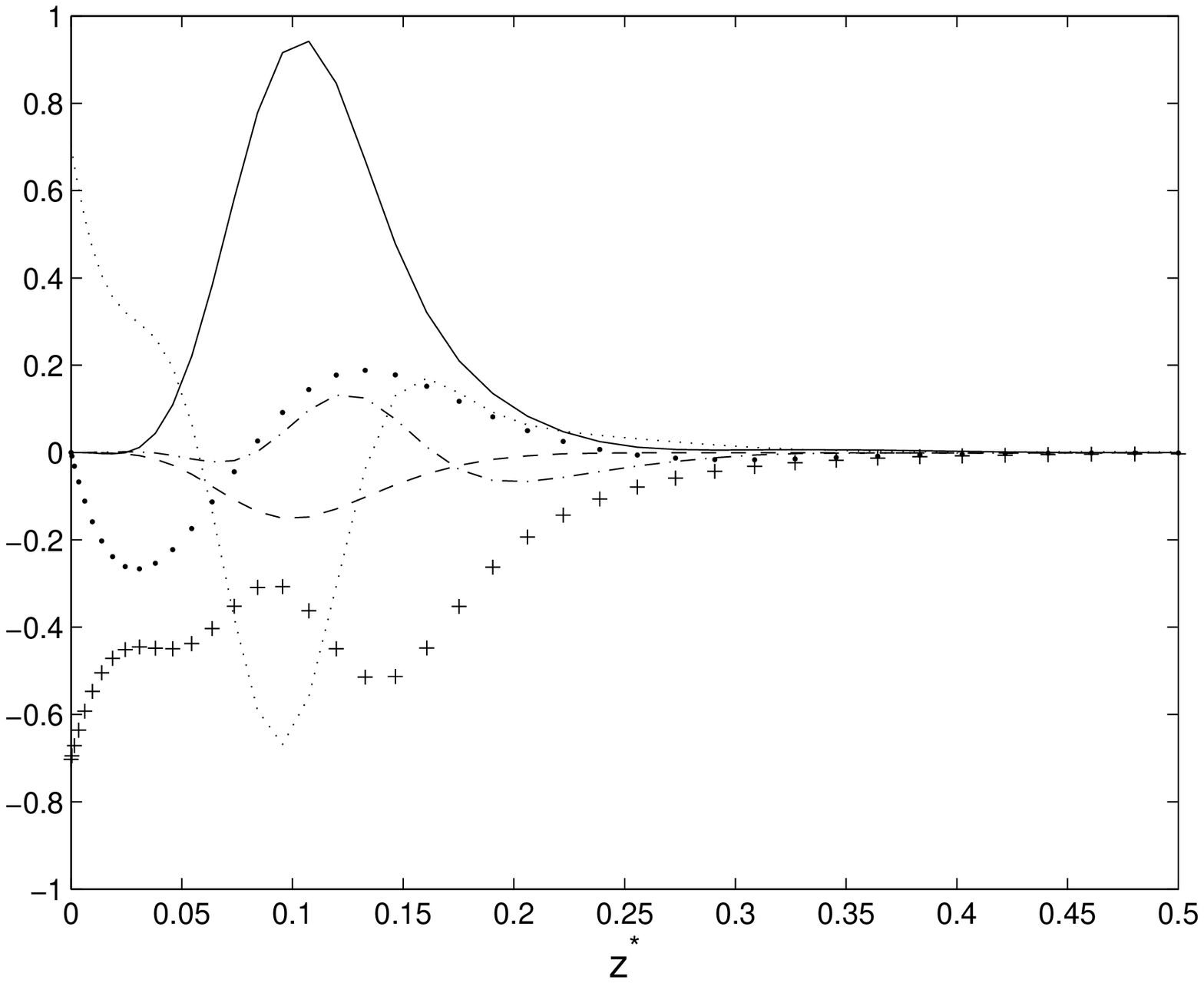}
\\
(c)
\end{tabular}
\\
\caption{Randriamampianina et Poncet, Phys. Fluids.} \label{bilankz}
\end{figure}

\newpage
\begin{figure}[!ht]
\centering \vspace{2cm}
\begin{picture}(280,180)
\put(-60,120){\shortstack{\large $z^*$}}
\put(-50,80){\shortstack{\large $1$}}
\put(-50,155){\shortstack{\large $0$}}
\put(-35,82){\includegraphics[width=0.187cm]{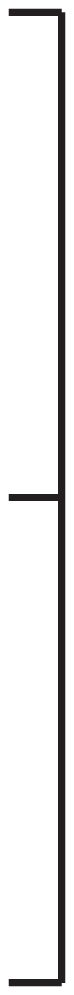}}
\put(-30,170){\includegraphics[angle=270,width=13.5cm]{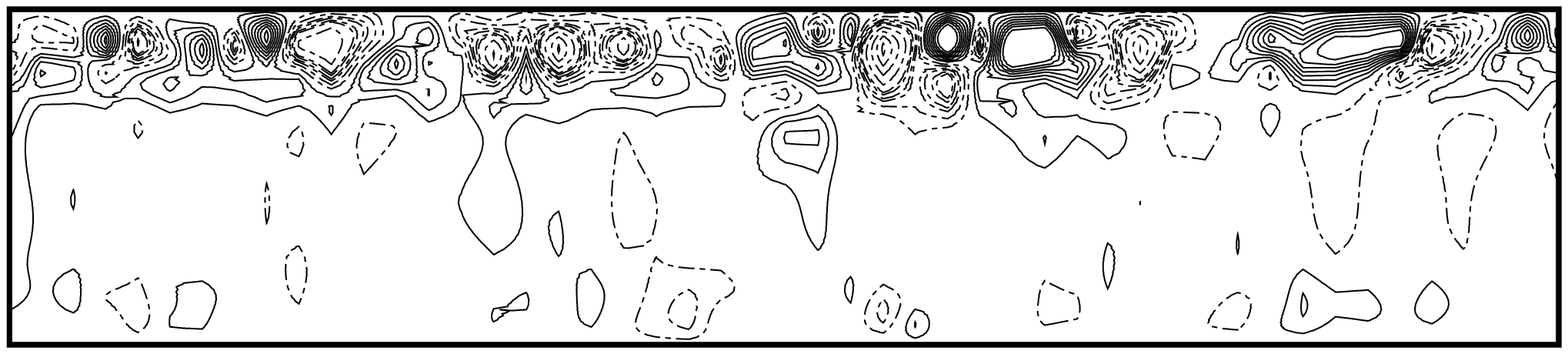}}
\put(145,170){\shortstack{\Large $v'_{\theta}$}}
\end{picture}
\begin{picture}(280,180)
\put(-60,120){\shortstack{\large $z^*$}}
\put(-50,80){\shortstack{\large $1$}}
\put(-50,155){\shortstack{\large $0$}}
\put(-35,82){\includegraphics[width=0.187cm]{REGLEY.eps}}
\put(-30,170){\includegraphics[angle=270,width=13.5cm]{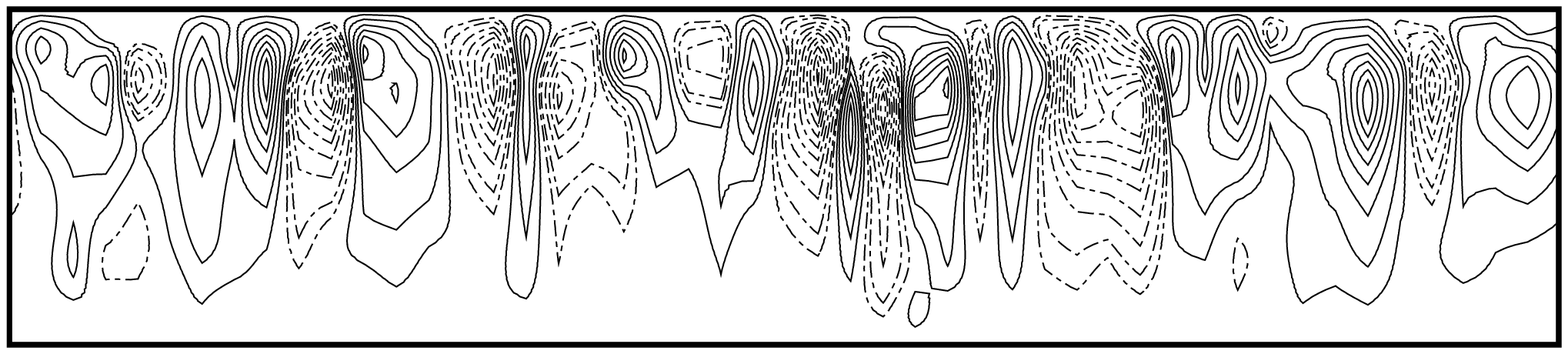}}
\put(145,170){\shortstack{\Large $v'_z$}}
\end{picture}
\begin{picture}(280,180)
\put(-60,120){\shortstack{\large $z^*$}}
\put(-50,80){\shortstack{\large $1$}}
\put(-50,155){\shortstack{\large $0$}}
\put(-35,82){\includegraphics[width=0.187cm]{REGLEY.eps}}
\put(-30,170){\includegraphics[angle=270,width=13.5cm]{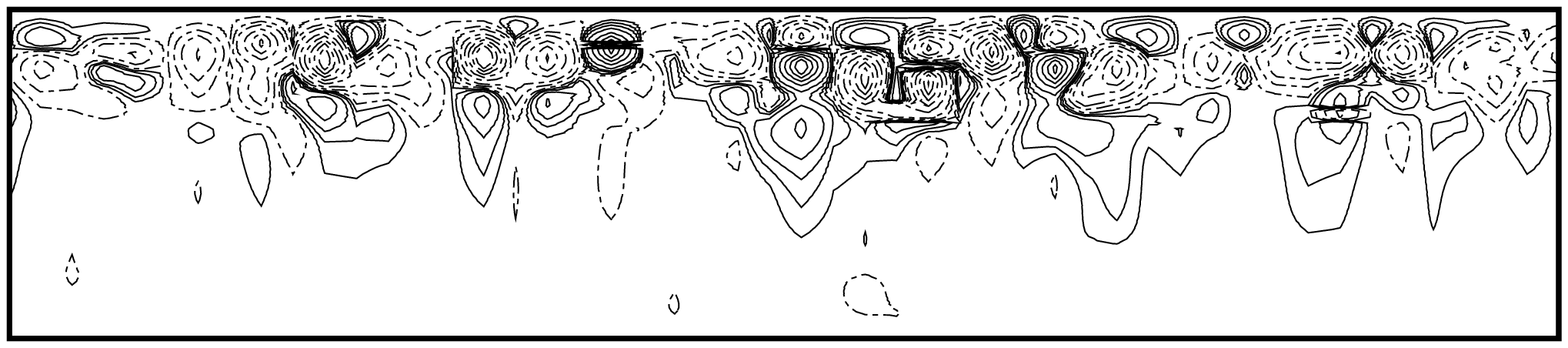}}
\put(140,170){\shortstack{\Large $-v'_{\theta}v'_z$}} \put(-31,75)
{\includegraphics[width=12.87cm]{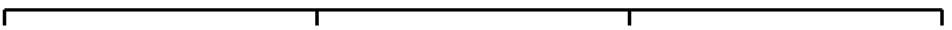}}
\put(-33,59){\shortstack{\large $0$}} \put(90,59){\shortstack{\large
$5$}} \put(208,59){\shortstack{\large $10$}}
\put(326,59){\shortstack{\large $15$}}
\put(140,59){\shortstack{\large $\Omega t$}}
\end{picture}
\caption{Randriamampianina et Poncet, Phys. Fluids.} \label{ztim80}
\end{figure}

\newpage
\begin{figure}[!ht]
\centering
\vspace{3cm}
\begin{tabular}{cc}
\includegraphics[width=7.5cm]{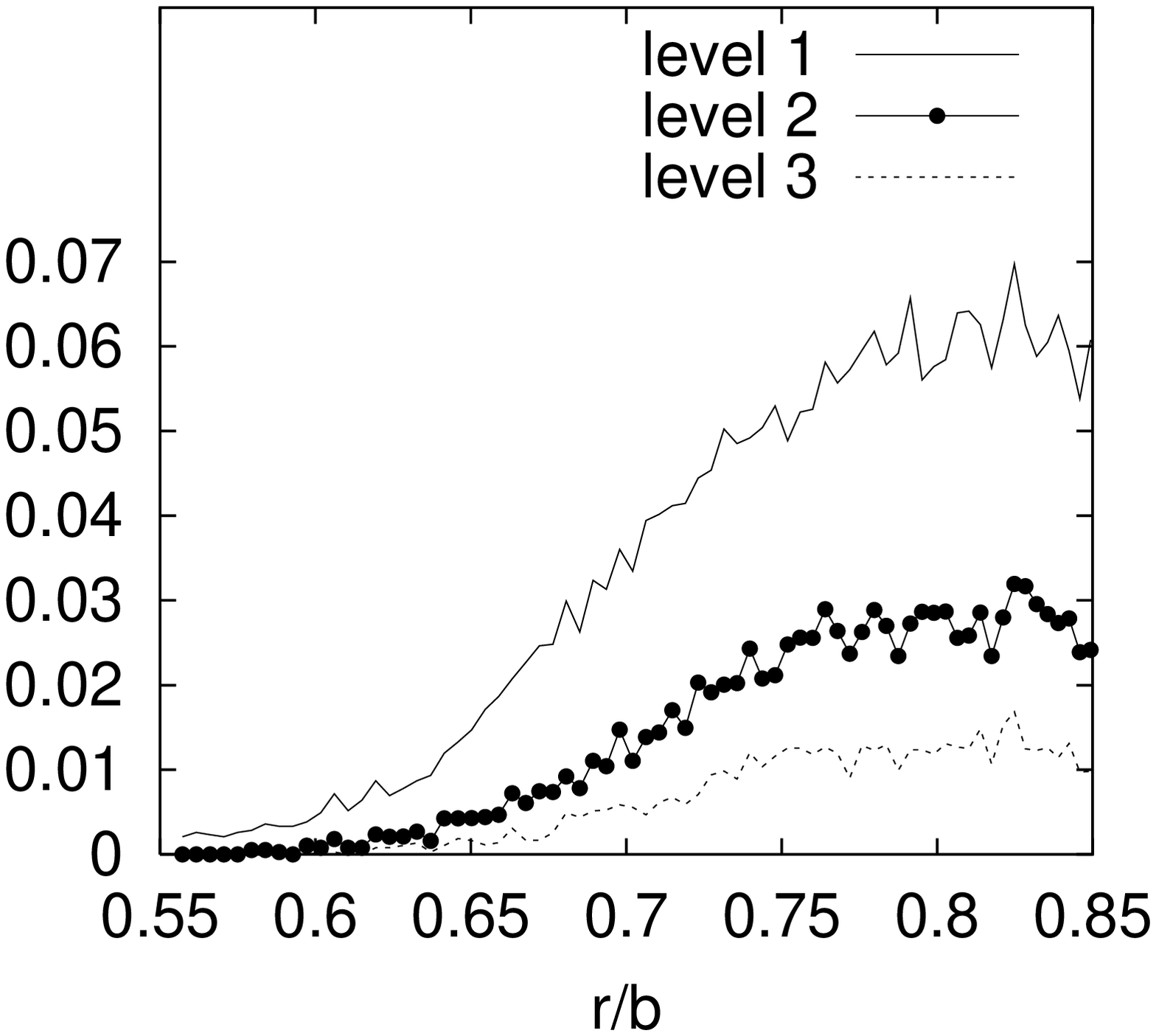} &
\includegraphics[width=7.5cm]{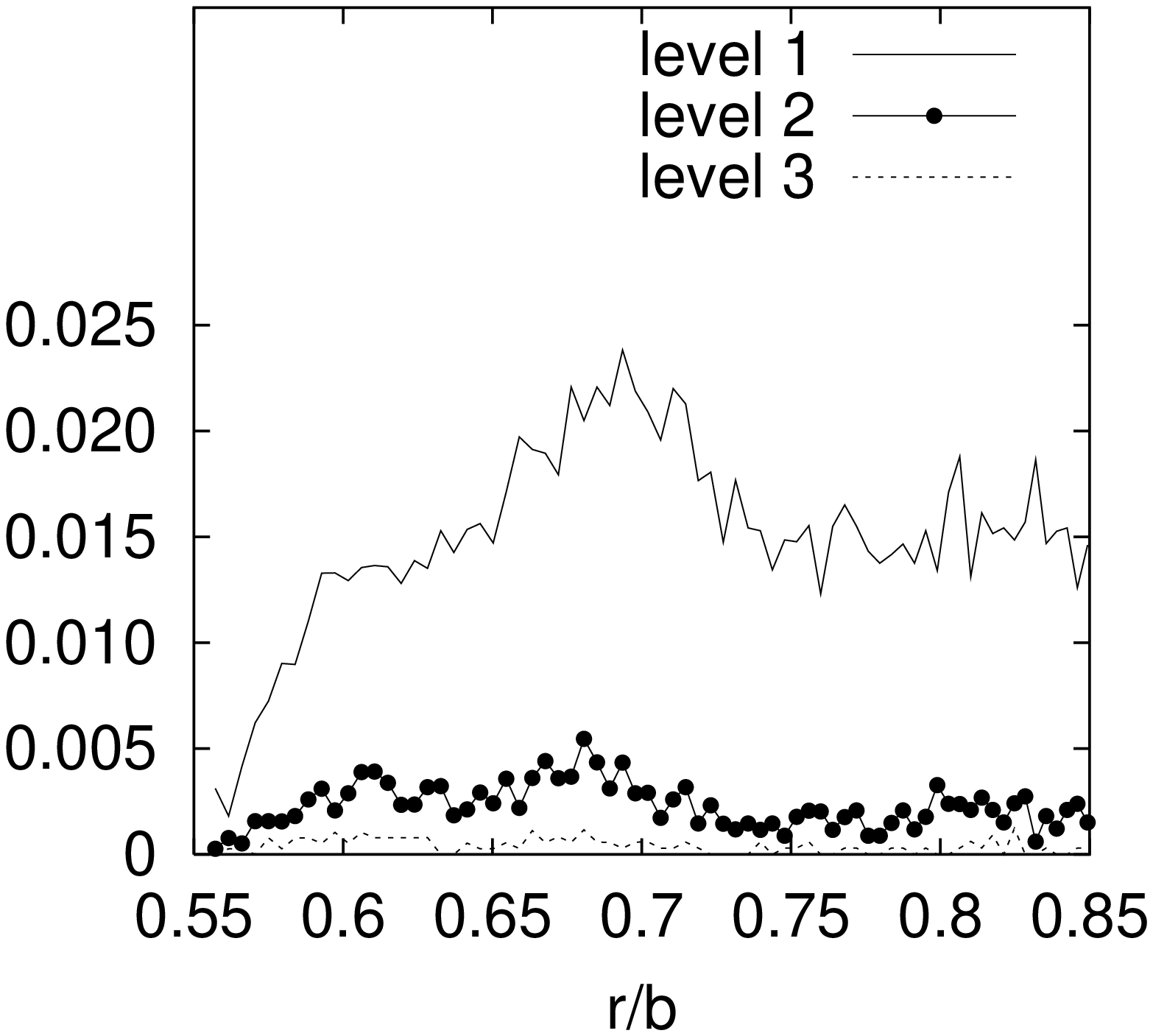}
\\
(a) & (b) \vspace{3cm}
\end{tabular}
\begin{tabular}{c}
\includegraphics[width=7.5cm]{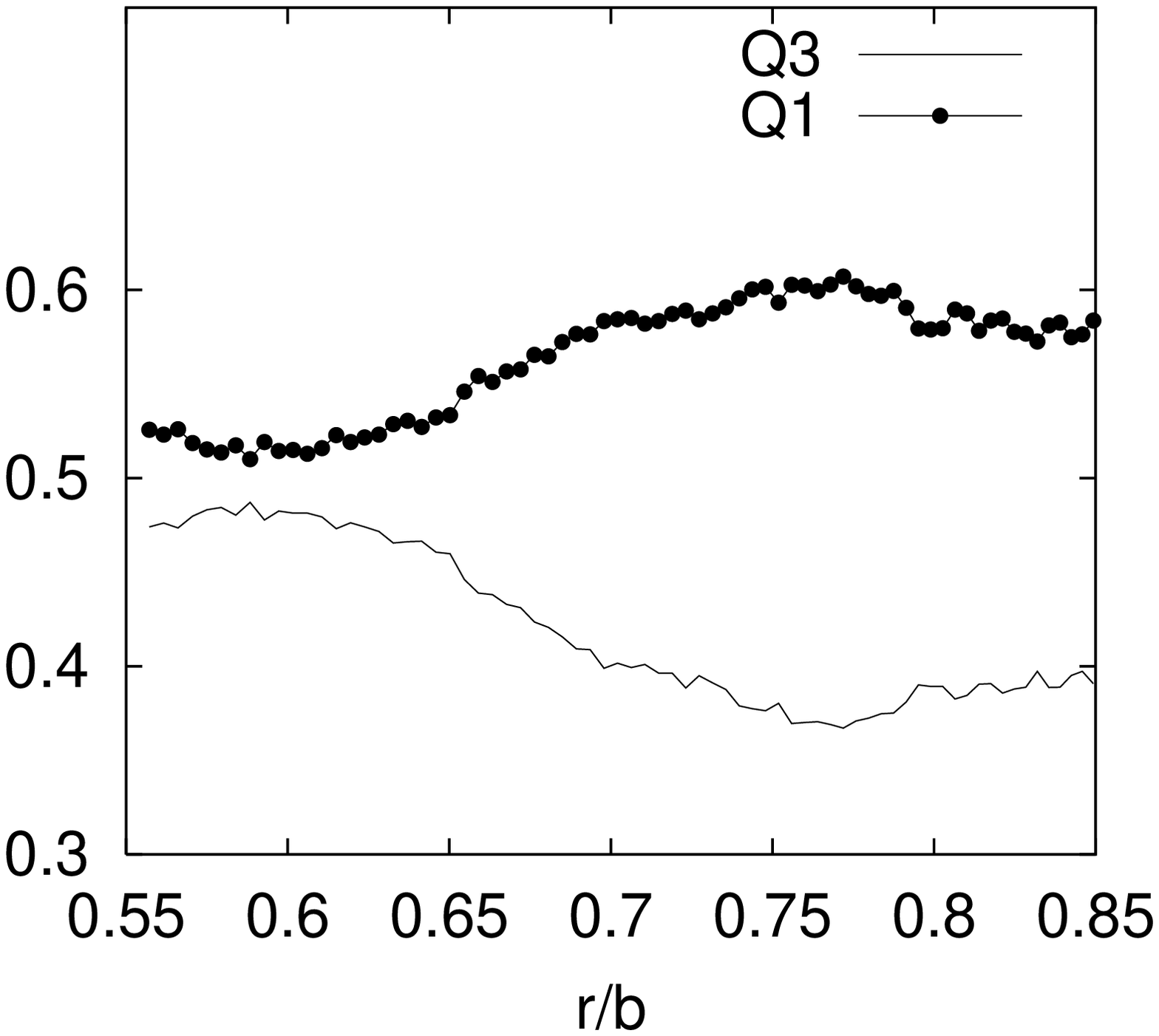}
\\
(c)
\end{tabular}
\\
\caption{Randriamampianina et Poncet, Phys. Fluids.} \label{percent}
\end{figure}

\newpage
\begin{figure}[!ht]
\centering
\vspace{3cm}
\begin{picture}(280,180)
\put(-50,190){\shortstack{ $<v'_{\theta}v'_z \mid strong$
$ejection>/<v'_{\theta}v'_z>$}} \put(165,190){\shortstack{
$<v'_{\theta}v'_z \mid strong$ $sweep>/<v'_{\theta}v'_z>$}}
\put(-60,3){\includegraphics[width=7cm]{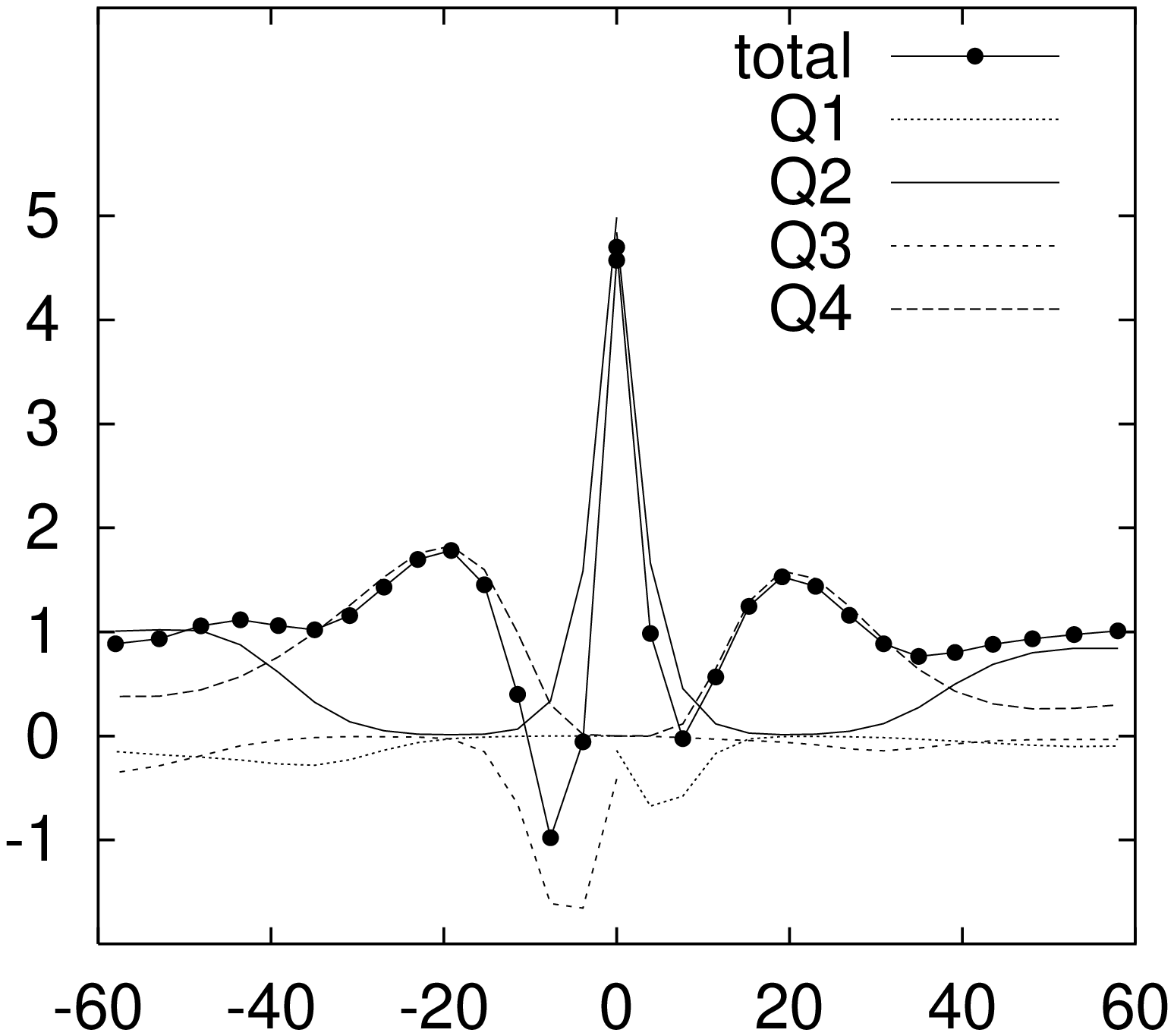}
\includegraphics[width=7cm]{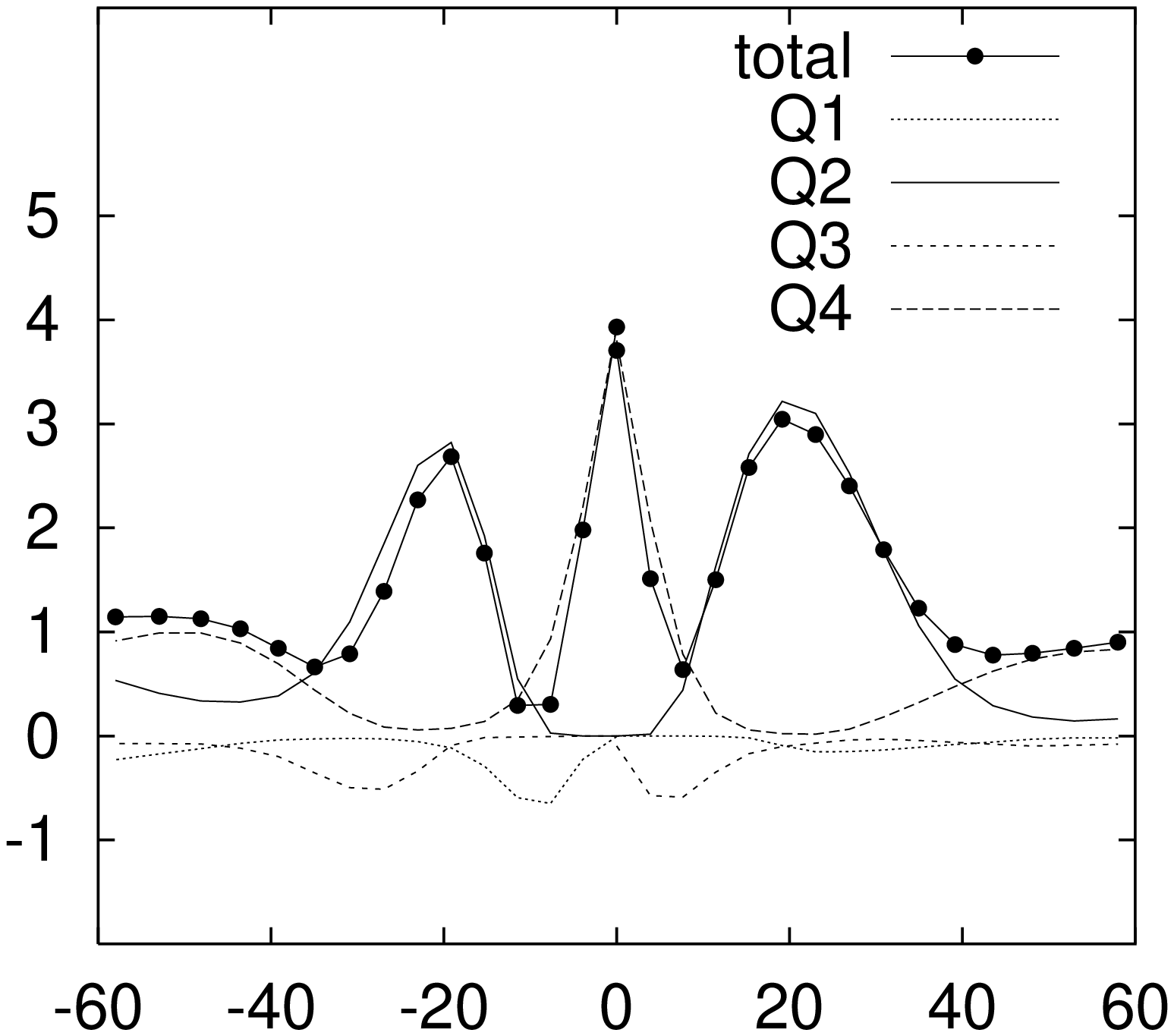}}
\put(30,-10){\shortstack{\large $-\Delta r^+$}}
\put(230,-10){\shortstack{\large $-\Delta r^+$}}
\end{picture}
\vspace{1cm}
\begin{tabular}{cccc}
\\
\hspace{2cm} & $(a)$ & \hspace{6cm} & $(b)$
\end{tabular}
\\
\caption{Randriamampianina et Poncet, Phys. Fluids.} \label{vwevent}
\end{figure}

\newpage
\begin{figure}[!ht]
\centering
\vspace{3cm}
\begin{picture}(280,180)
\put(-50,190){\shortstack{ $<v'_{\theta} \mid strong$
$ejection>/<\sqrt{\overline{v'^2_{\theta}}}>$}}
\put(165,190){\shortstack{$<v'_z \mid strong$
$ejection>/<\sqrt{\overline{v'^2_z}}>$}}
\put(-60,3){\includegraphics[width=7cm]{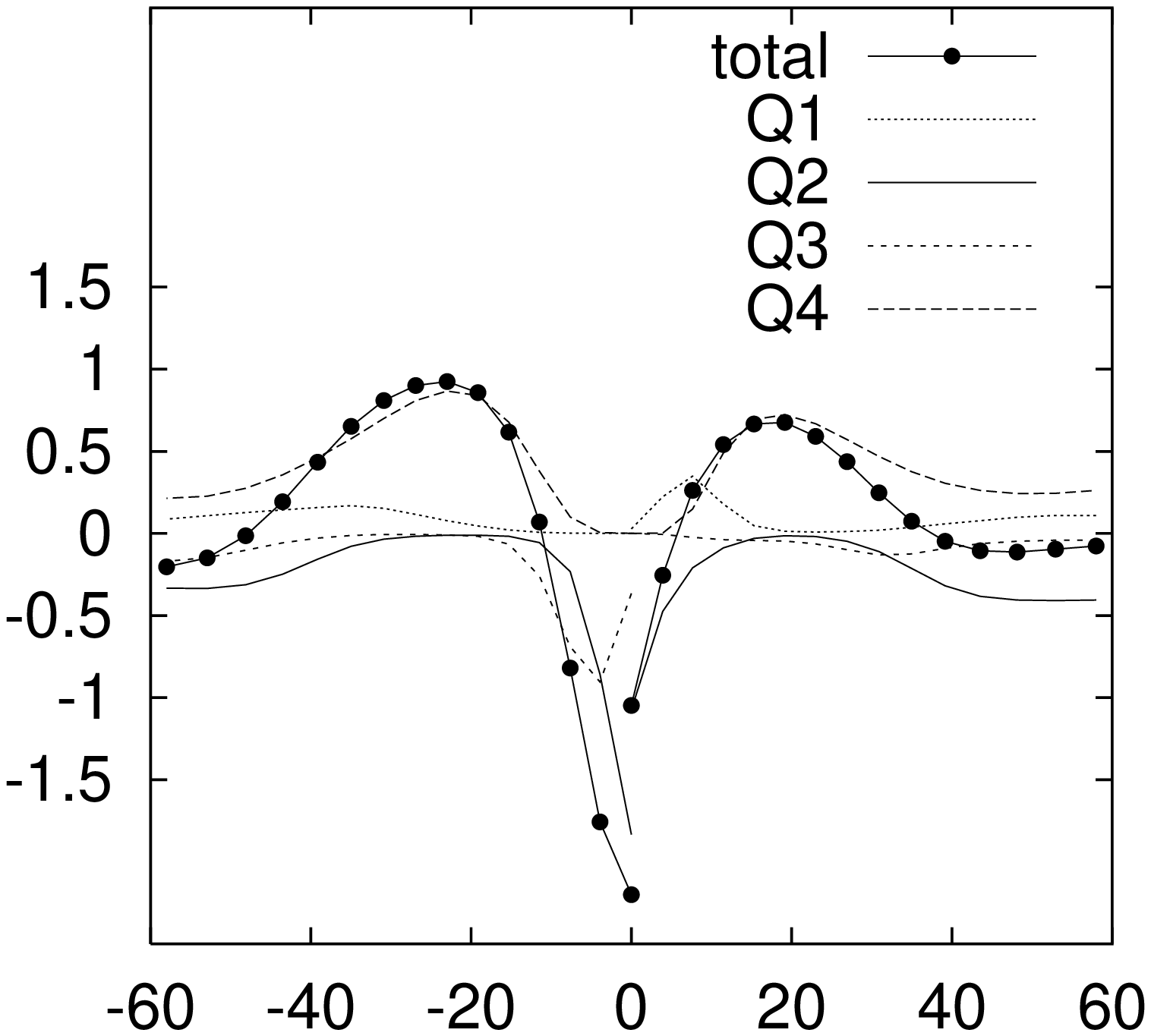}
\includegraphics[width=7cm]{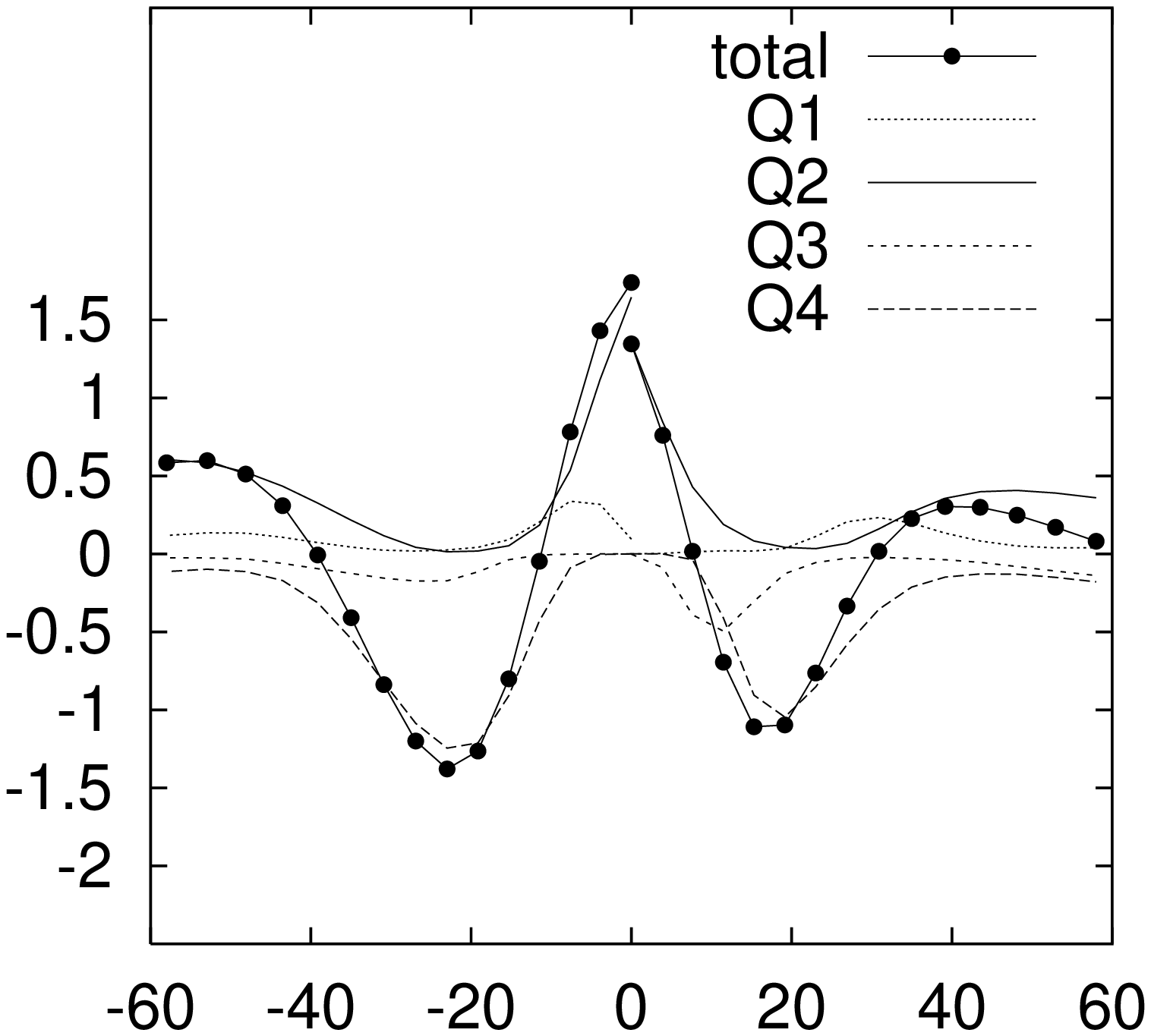}}
\put(30,-10){\shortstack{\large $-\Delta r^+$}}
\put(230,-10){\shortstack{\large $-\Delta r^+$}}
\end{picture}
\vspace{1cm}
\begin{tabular}{cccc}
\\
\hspace{2cm} & $(a)$ & \hspace{6cm} & $(b)$
\end{tabular}
\\
\caption{Randriamampianina et Poncet, Phys. Fluids.} \label{vweject}
\end{figure}

\newpage
\begin{figure}[!ht]
\centering
\vspace{3cm}
\begin{picture}(280,180)
\put(-50,190){\shortstack{ $<v'_{\theta} \mid strong$
$sweep>/<\sqrt{\overline{v'^2_{\theta}}}>$}}
\put(165,190){\shortstack{$<v'_z \mid strong$
$sweep>/<\sqrt{\overline{v'^2_z}}>$}}
\put(-60,3){\includegraphics[width=7cm]{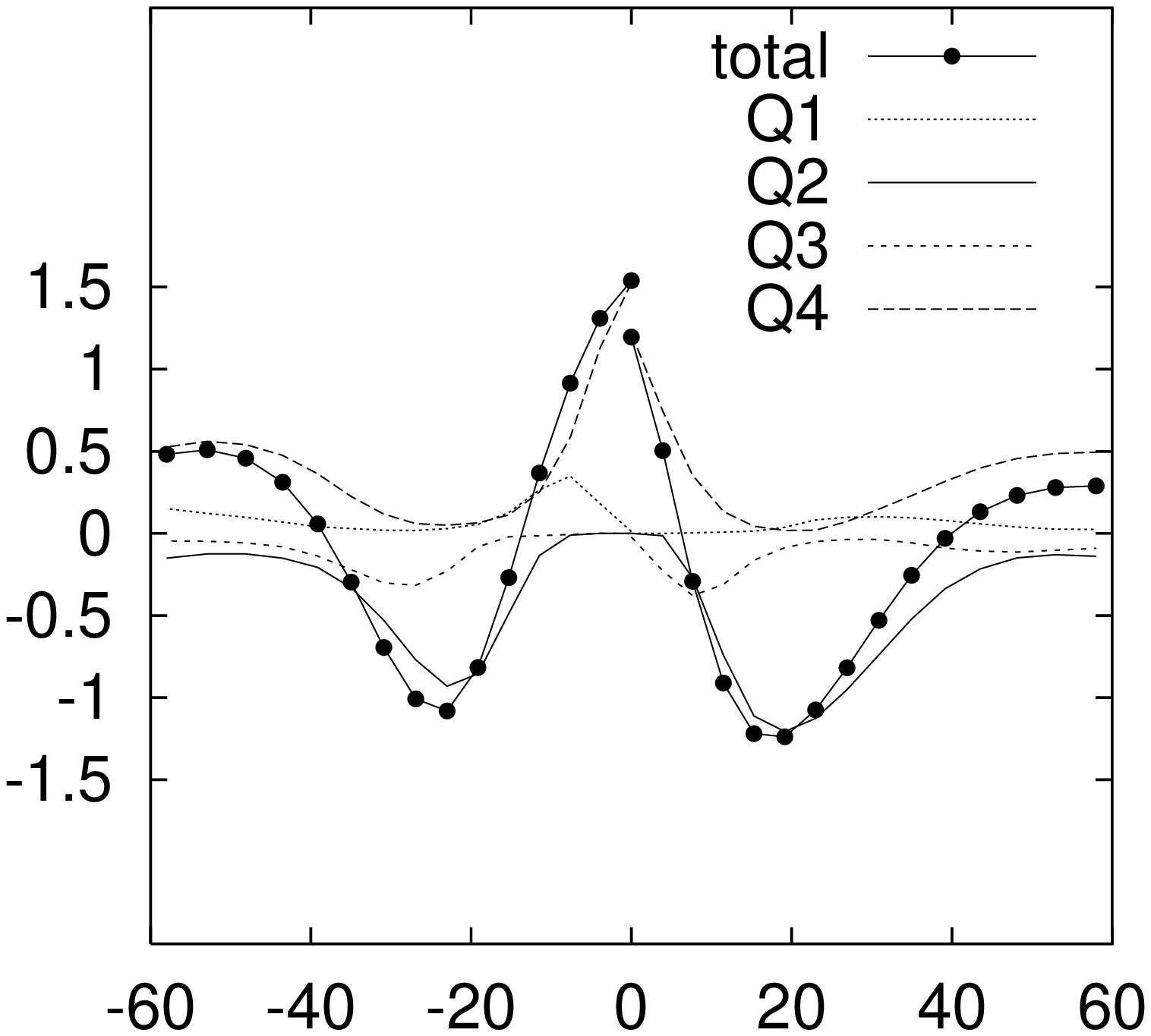}
\includegraphics[width=7cm]{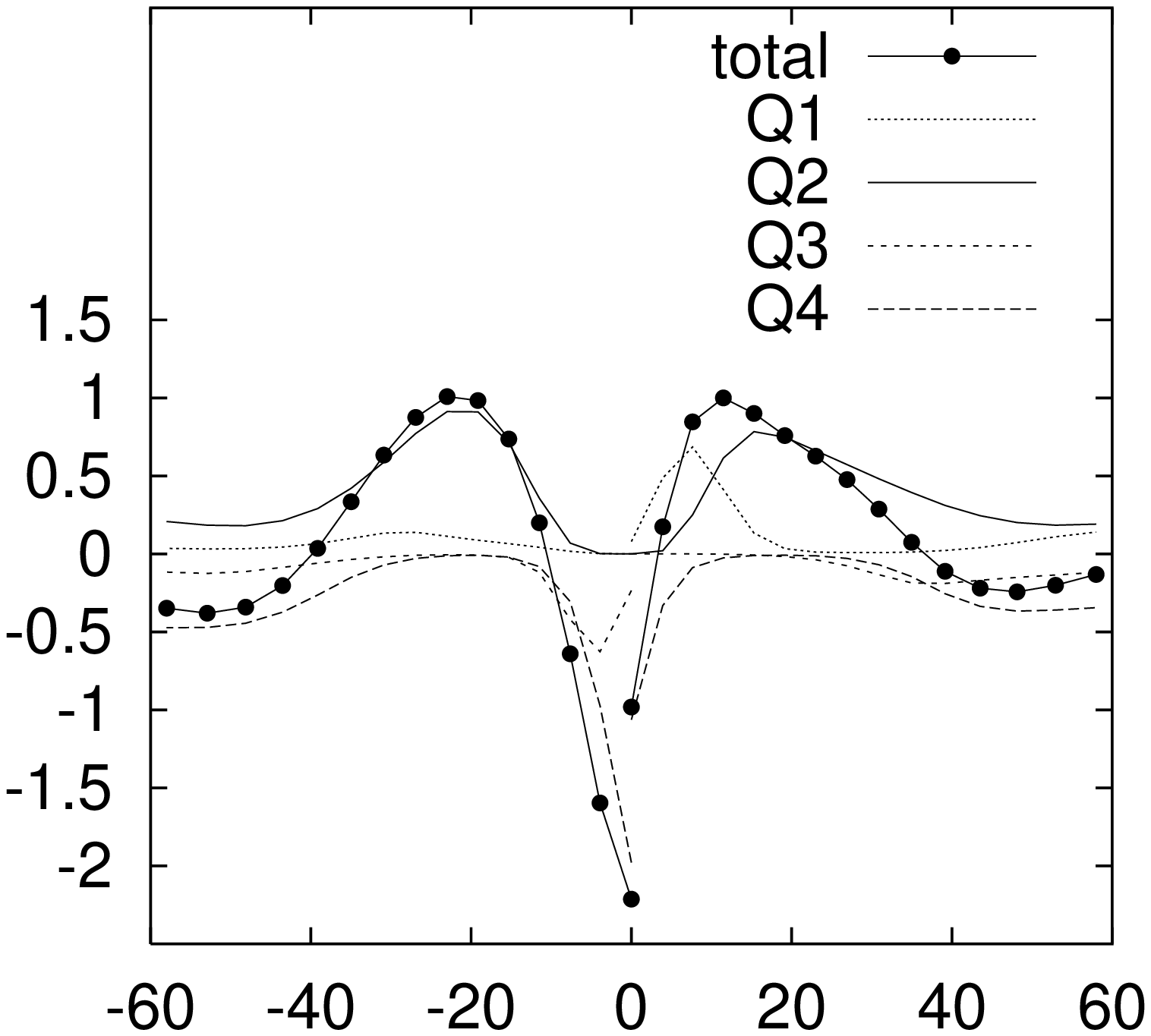}}
\put(30,-10){\shortstack{\large $-\Delta r^+$}}
\put(230,-10){\shortstack{\large $-\Delta r^+$}}
\end{picture}
\vspace{1cm}
\begin{tabular}{cccc}
\\
\hspace{2cm} & $(a)$ & \hspace{6cm} & $(b)$
\end{tabular}
\\
\caption{Randriamampianina et Poncet, Phys. Fluids.} \label{vwsweep}
\end{figure}

\clearpage

\pagestyle{empty}
\newpage
\appendix
\section{Balance equations for the kinetic energy budgets in cylindrical coordinates}

\begin{eqnarray} \label{TKEBA}
\overline{V_j}k_{,j}=-R_{ij}\overline{V_{i,j}}-\frac{1}{2}\overline{(v'_mv'_jv'_j)_{,m}}
-\frac{1}{\rho}\overline{(v'_jp')_{,j}}-\nu\overline{v'_{j,m}v'_{j,m}}+\nu
k_{,jj}
\end{eqnarray}
\[(a) \qquad \qquad (b) \qquad \qquad  (c) \qquad \qquad (d)
\qquad \qquad (e) \qquad \qquad (f)
\]
Each instantaneous variable is decomposed in the following form:
\[
f = \overline{f} + f'
\]
where $\overline{f}$ is the time averaged component and $f'$ the fluctuating part.
Using the normalizing scale $2 \Omega^3
h^2$ leads to the following dimensionless terms:
\begin{eqnarray}
(a) \equiv G \overline{V}_r\partial_rk +\frac{G}{(r+R_c)}
 \overline{V}_{\theta}\partial_{\theta}k
 + \overline{V}_z\partial_zk
\end{eqnarray}
\begin{eqnarray}
(b) \equiv G R_{rr}\partial_r{\overline{V}_r} +
R_{zz}\partial_z{\overline{V}_z}+ G R_{rz}
(\partial_r{\overline{V}_z} + \partial_z{\overline{V}_r})
+ \nonumber \\
G R_{r\theta}
(\partial_r\overline{V}_{\theta}+\frac{1}{(r+R_c)}\partial_{\theta}\overline{V}_r
- \frac{\overline{V}_{\theta}}{r+R_c}) + \nonumber \\
R_{z\theta}(\partial_z\overline{V}_\theta+\frac{\partial_{\theta}\overline{V}_z}{r+R_c})
+ G R_{\theta\theta}(\frac{\overline{V}_r}{r+R_c} +
\frac{\partial_{\theta}\overline{V}_{\theta}}{r+R_c})
\end{eqnarray}
\begin{eqnarray}
(c) \equiv G \partial_r\overline{(v'_rk')} +
\partial_z\overline{(v'_zk')} + G \frac{\overline{v'_rk'}}{r+R_c} +\frac{G}{r+R_c}\partial_{\theta}\overline{(v'_{\theta}k')}
\end{eqnarray}
\begin{eqnarray}
(d) \equiv G \partial_r\overline{(v'_rp')} +
\partial_z\overline{(v'_zp')}
 + G \frac{\overline{v'_rp'}}{r+R_c} +\frac{G}{r+R_c}\partial_{\theta}\overline{(v'_{\theta}p')}
\end{eqnarray}
\begin{eqnarray}
(e) \equiv \frac {R_c +1}{G Re}[G^2
\overline{\partial_rv'_r\partial_rv'_r}+\overline{\partial_zv'_z\partial_zv'_z}+
G^2 \overline{\partial_ru'_z\partial_rv'_z} +
\overline{\partial_zv'_r\partial_zv'_r} + \nonumber \\
G^2 \overline{\partial_rv'_\theta\partial_rv'_\theta} +
\overline{\partial_zv'_\theta\partial_zv'_\theta}+ G^2
 \overline{(\frac{1}{r+R_c}\partial_{\theta}v'_{\theta}+\frac{v'_r}{r+R_c})^2} +
 \nonumber \\
 G^2 \overline{(\frac{1}{r+R_c}\partial_{\theta}v'_r-\frac{v'_\theta}{r+R_c})^2}
+ G^2
\overline{\frac{\partial_{\theta}v'_{\theta}\partial_{\theta}v'_{\theta}}{(r+R_c)^2}}]
\end{eqnarray}
\begin{eqnarray}
(f) \equiv \frac {R_c +1}{G Re}(G^2 \partial_{rr}k +
 \partial_{zz}k
 + G^2 \frac{\partial_r k}{r+R_c}+
\frac{G^2}{(r+R_c)^2}\partial_{{\theta}{\theta}}k)
\end{eqnarray}
where
\[
 k'=\frac{v'^2_r+v'^2_\theta+v'^2_z}{2},\qquad
k=\overline{\frac{v'^2_r+v'^2_\theta+v'^2_z}{2}} =
\overline{k'},\qquad R_{ij}=\overline{v'_iv'_j}.
\]
\end{document}